\documentclass[11pt,a4paper]{article}
\usepackage{heppub}
\pdfoutput=1
\usepackage{braket, slashed, placeins,multirow,colortbl,makecell,subcaption,mathtools, slashed}
\usepackage[svgnames,dvipsnames]{xcolor}
\usepackage[T1]{fontenc}
\usepackage[utf8]{inputenc}
\usepackage{dsfont}
\usepackage{tabularx}

\newcommand{\cC}{\ensuremath{\mathcal{C}}}
\newcommand{\cCK}{\ensuremath{\mathcal{CK}}}

\newcommand{\cH}{\ensuremath{\mathcal{H}}}
\newcommand{\cK}{\ensuremath{\mathcal{K}}}
\newcommand{\cL}{\ensuremath{\mathcal{L}}}

\newcommand{\cO}{\ensuremath{\mathcal{O}}}
\newcommand{\cP}{\ensuremath{\mathcal{P}}}
\newcommand{\cPT}{\ensuremath{\mathcal{PT}}}
\newcommand{\cT}{\ensuremath{\mathcal{T}}}
\newcommand{\Tr}{\text{Tr}}  
\newcommand{\zn}{\ensuremath{\mathbb{Z}_N}} 
\newcommand{\zz}{\ensuremath{\mathbb{Z}_3}} 
\newcommand{\mub}{\ensuremath{\mu_{\text{\scriptsize B}}}}
\newcommand{\psibar}{\bar{\psi}}
\newcommand{\bR}{\ensuremath{\mathbb{R}}}
\newcommand{\bZ}{\ensuremath{\mathbb{Z}}}
\newcommand{\cZ}{\ensuremath{\mathbb{Z}}}

\DeclareRobustCommand{\change}[1]{{\color{black} #1}}

\begin{document}
	\title{Exotic phases in finite-density $\mathbb Z_3$ theories}
	\author[a]{Michael C. Ogilvie,}
	\author[a]{ Moses A. Schindler,}
	\author[b,c]{ and Stella T.~Schindler}
	\affiliation[a]{Physics Department, Washington University in St. Louis\\
		1 Brookings Drive, St. Louis, MO 63130, USA}
	\affiliation[b]{Theoretical Division, Los Alamos National Laboratory\\
		Los Alamos, NM 87545, USA}
	\affiliation[c]{Center for Theoretical Physics, Massachusetts Institute of Technology\\
		77 Massachusetts Ave., Cambridge, MA 02139, USA}
	\emailAdd{mco@wustl.edu}
	\emailAdd{schindler@lanl.gov}
	
\abstract{
Lattice $\zz$ theories with complex actions share many key features with finite-density QCD including a sign problem and $\cCK$ symmetry. Complex $\zz$ spin and gauge models exhibit a generalized Kramers-Wannier duality mapping them onto chiral $\zz$ spin and gauge models, which are simulatable with standard lattice methods in large regions of parameter space. The Migdal-Kadanoff real-space renormalization group (RG) preserves this duality, and we use it to compute the approximate phase diagram of both spin and gauge $\zz$ models in dimensions one through four. Chiral $\zz$ spin models are known to exhibit a Devil's Flower phase structure, with inhomogeneous phases that can be thought of as $\zz$ analogues of chiral spirals. Out of the large class of models we study, we find that only chiral spin models and their duals have a Devil's Flower structure with an infinite set of inhomogeneous phases, a result we attribute to Elitzur's theorem. We also find that different forms of the Migdal-Kadanoff RG produce different numbers of phases, a violation of the expectation for universal behavior from a real-space RG. We discuss extensions of our work to $\zn$ models, SU($N$) models and nonzero temperature.}
	
	\preprint{\vbox{\hbox{MIT-CTP 5795, LA-UR-24-31042}}}
	\keywords{}\arxivnumber{}	\maketitle
	\allowdisplaybreaks
	\sloppy
	
\section{Introduction}\label{sec:intro}

A central objective of nuclear physics is mapping out the phases of QCD in the baryon density-temperature  ($\mub$-$T$) plane \cite{796947, Akiba:2015jwa, Achenbach:2023pba}.  Several major collider experiments are pursuing this goal, including RHIC at Brookhaven \cite{STAR:2002eio, PHENIX:2015siv}, ALICE at CERN \cite{ALICE:2008ngc}, and the planned CBM experiment at FAIR \cite{CBM:2016kpk}. To complement this work, a simultaneous theoretical effort is targeting QCD phase structure from first principles \cite{Busza:2018rrf}; however, at present we lack rigorous and systematically improvable methods to probe much of $\mub$-$T$ plane. On the numerical front, lattice QCD calculations at $\mub \gtrsim T$ are obstructed by a sign problem \cite{Aarts:2015tyj,Nagata:2021ugx, Ratti:2018ksb, Ratti:2021ubw}. While many approaches are under development to overcome sign problems \cite{Barbour:1997ej, Fodor:2001au, Fodor:2001pe, Fodor:2007vv, Alexandru:2014hga,Alexandru:2005ix, Kratochvila:2005mk, Ejiri:2008xt,Gattringer:2014nxa,Cristoforetti:2012su, Alexandru:2015xva, Alexandru:2015sua,Parisi:1983mgm, Attanasio:2020spv,Albergo:2019eim, Lawrence:2021izu,Rico:2013qya, Pichler:2015yqa, Kadoh:2018hqq, Kadoh:2018tis, Kuramashi:2018mmi, Magnifico:2020bqt, Yuan:2020xmq, Meurice:2022xbk}, they cannot fully handle QCD yet. On the analytic front, various effective field theories (EFTs) have been developed such as hard thermal loops \cite{Braaten:1989mz, Andersen:2004fp} and high density effective theory \cite{Hong:1999ru, Schafer:2003jn}, but these only cover limited regions of parameter space. In the near term, there exist few viable first-principles approaches to finite-density QCD, such as Taylor expansions off the $T$-axis \cite{Allton:2002zi, Allton:2005gk, Gavai:2008zr, Kaczmarek:2011zz} and studies at imaginary baryon chemical potential \cite{Dagotto:1989fw, Hasenfratz:1991ax, Alford:1998sd, deForcrand:2002hgr, DElia:2002tig}.  

A more recently proposed approach to understanding finite-density QCD capitalizes on one of its symmetries, combined charge and complex conjugation symmetry $\cCK$ \cite{Meisinger:2012va, Schindler:2019ugo, Schindler:2021otf}. We can see this symmetry directly from the QCD Lagrangian at finite density \cite{Schindler:2021otf}, 
\begin{align}\label{eq:euclidean-qcd}
	\cL_{\rm QCD}^{\rm Eucl} &= \bar\psi_j (i\slashed{D}-m)\psi_j+\frac{1}{4}G_{\mu\nu}^AG_{\mu\nu A} + i\mub \bar\psi \gamma^4 \psi\,,
\end{align}
where $\psi_j$ are fermion fields, $D$ is a covariant derivative, $G$ is the gluon field strength, $\mub$ is the baryon chemical potential, and $\gamma^i$ represents the standard Dirac matrices. 
It is well known that when $\mub=0$, QCD preserves charge conjugation $\cC$, parity inversion $\cP$, and time-reversal $\cT$ symmetries individually and in combination. The situation changes substantially when we turn on a nonzero baryon density term $\mub$ in \eq{euclidean-qcd}. While the fermion bilinear $\bar\psi \gamma^4 \psi$ remains invariant under $\cP$ and $\cT$, it flips sign under $\cC$.  In turn, the lattice path integral weights in \eq{euclidean-qcd} and the fermion determinant become complex, the transfer matrix becomes non-Hermitian, and QCD develops a sign problem. Additionally, \eq{euclidean-qcd} loses Lorentz invariance at $\mub \neq 0$, as the chemical potential picks out a preferred direction ($\gamma^4$ rather than $\gamma^\mu$). However,  the complex conjugation operation $\cK$ flips the sign of $i$ while leaving the constant $\mub$ and the bilinear $\psibar\gamma^4\psi$ invariant. Thus, the combined operation $\cCK$ leaves the entire Lagrangian invariant.

The operation $\cCK$ belongs to the class of $\cPT$-type symmetries, which are widely studied in optics and condensed matter for their unique properties and extensive experimental applications \cite{Bender:1998ke,Bender:2019cwm,Bender:2023cem,RevModPhys.88.035002, Feng2017, Ozdemir2019, Miri2019, Ashida:2020dkc, Ding:2022juv}. 
A $\cPT$-type symmetry is any symmetry combining one linear operator (e.g. $\cP$ or $\cC$) and one  antilinear operator (e.g. $\cT$ or $\cK$). 
Importantly, $\cPT$-symmetric matrices may be non-Hermitian; however, every eigenvalue of a $\cPT$-symmetric system is always either real or part of a complex-conjugate pair \cite{Bender:1998ke, Bender:2007nj}. 
The study of quantum field theories with non-Hermitian transfer matrices and $\cPT$-type symmetries is a far younger field than $\cPT$-symmetric quantum mechanics and optics. Nonetheless, many steps have been taken towards developing a formal understanding of and techniques for these systems  \cite{Bender:2003ve, Bender:2004sv, Bender:2007wu,Ivanov:2007me, Curtright:2006aq, Curtright:2007wh, Ai:2022csx, Lawrence:2023woz,  Romatschke:2022llf, Bender:2018pbv, Bender:2022eze, Alexandre:2017foi, Felski:2021evi, Bender:2021fxa, Fring:2020wrj, Fring:2021zci, Correa:2021pwi, Chernodub:2017lmx}. A number of $\cPT$-QFTs have been studied within the context of Beyond the Standard Model (BSM) model building \cite{Jones-Smith:2009qeu, Ohlsson:2015xsa, Chen:2024bya, Alexandre:2015kra, Alexandre:2020tba, Mavromatos:2020hfy, Alexandre:2020bet, Mavromatos:2021hpe, Mavromatos:2022heh, Ogilvie:2021wvb, Mannheim:2011ds}. Within the Standard Model, several models sharing features of dense QCD have been analyzed using tools based on non-Hermiticity and $\cCK$ symmetry \cite{Meisinger:2012va, Nishimura:2014rxa, Nishimura:2014kla, Nishimura:2015lit, Nishimura:2016yue, Schindler:2021cke,Meisinger:2013zba}.

Most importantly for this work, theories with non-Hermitian transfer matrices and $\cPT$-type symmetries generally support unusual phase structure not seen in conventional field theories \cite{Schindler:2019ugo, Schindler:2021cke, Schindler:2021otf, Ogilvie:2021wvb}. The appearance of complex conjugate pairs of transfer matrix eigenvalues is a hallmark of $\cPT$-type symmetries, leading to sinusoidally-modulated exponential decay in correlation functions, similar to Friedel oscillations \cite{fetter2012quantum, Kapusta:1988fi}. Regions of parameter space with conjugate pairs may also behave as \textit{moat regimes} \cite{Pisarski:2021qof}, a name that originates in the condensed matter literature \cite{Pu2024,Sedrakyan14}.
The boundary between a region where all eigenvalues are positive and a region with complex conjugate pairs is referred to in the statistical mechanics literature as a \textit{disorder line} \cite{stephenson_ising_1970}. $\cPT$-type symmetric transfer matrices may also lead to inhomogeneous phases as a consequence of a Lifshitz instability. We summarize the possible behaviors in \tab{mass-matrix}.
Note that an arbitrary non-Hermitian system without a $\cPT$-type symmetry generally has a mix of positive, negative, and complex eigenvalues and thus does not in general support the types of stable exotic phases we see in $\cPT$ systems. 

\begin{table}\def\arraystretch{1.3}
	\begin{center}
		\begin{tabular}{|c|c|}
			\hline
			\rowcolor{LightCyan}{\bf Transfer matrix eigenvalues} & {\bf Phase behavior}
			\\ \hline
			All positive & Normal 
			\\ \hline
			Some complex conjugate pairs & Complex (Friedel-like) 
			\\ \hline
			\null \quad Even number of negative eigenvalues\quad\null  & \null\quad Inhomogeneous (Lifshitz instability) \quad\null
			\\ \hline
			Odd number of negative eigenvalues & Unstable
			\\ \hline
		\end{tabular}
	\end{center}
	\caption{Relationship between the phases in a moat regime and the spectral properties of a theory; see  \refcite{Schindler:2019ugo}. A disorder line marks the onset of a complex phase.}\label{tab:mass-matrix}
\end{table}

The loss of Hermiticity and/or Lorentz invariance can give rise to non-positivity, moat regimes, and their associated phases. A large body of literature on  inhomogeneous phases has been developed in condensed matter systems, which are naturally non-relativistic \cite{cholesterolstripes, kasper,nelson,sadoc,sethna,ref:nussinovAPT,glass_review,jorg,LC1,nematicma,LC2,LC3,Peierls,LL5,deGennes,Als-Nielsen, Kivelson1998,tran,steve,Low,saxena,stroud,kabanov,cuprate1, cuprate2,cuprate3, cuprate4,qhstr2,QHE,fogler1,chalker,qhstr1, qhstr3,magnstr1,magnstr2,magnetic3, magnstr3, magnstr4,magnstr5, magnstr6,magneticgarnet,highT, membranestripes, dipole1,vindigni, bates,leibler,rosedale, ciah, Gennady,  pnic1, pnic2, cheong,golosov,salamonmanganites, jan,rmp_steve, annni1,annni2,Selke-review, FK_model,devil, liebfrusferr,ortix,gulacsi,barci, Pokrovsky-Talapov, Braz, modulation2,modulation1, Mukamel}; these phases are often associated with competing interactions \cite{Seul,rc1,rc2,rc3,  Derrick, Leonid, compete}. 
In many models sharing features with QCD, Friedel-like phases and disorder lines have been observed, including in flux tube models \cite{Patel:2011dp}, finite-density Potts models \cite{Akerlund:2016myr}, PNJL models \cite{Nishimura:2014rxa, Nishimura:2014kla}, static quark models at strong coupling \cite{Nishimura:2015lit}, liquid-gas models \cite{Nishimura:2016yue}, mass-mixing models \cite{Schindler:2019ugo}, and heavy quark models \cite{Schindler:2021cke}. 
Inhomogeneous phases have also been explored in models sharing features with QCD, including $\cO(N)$ models  \cite{Nussinov:2001ix,nussinov2004commensurate,PhysRevE.86.041132,Pisarski:2020dnx, Winstel:2024qle}, scalar models \cite{Pisarski:2019cvo, Schindler:2019ugo}, various Gross-Neveu models \cite{Thies:2003kk,Basar:2008im, Basar:2008ki, Lenz:2020bxk, Buballa:2020nsi, Koenigstein:2021llr,Pannullo:2021edr, Lenz:2021kzo, Lenz:2021vdz, Lenz:2021kzo, Nonaka:2021pwm, Pannullo:2021edr, Winstel:2021yok, Winstel:2024dqu, Koenigstein:2024cyq}, NJL models \cite{Basar:2009fg, Carignano:2019ivp, Pannullo:2022eqh, Pannullo:2023cat, Koenigstein:2023yzv,  Motta:2023pks, Pannullo:2024sov},  PNJL models \cite{Carignano:2010ac}, Yukawa models \cite{Pannullo:2023one}, quark-meson models \cite{Carignano:2014jla, Buballa:2018hux, Buballa:2020xaa, Haensch:2023sig}, and functional renormalization group (FRG) studies of QCD \cite{Fu:2019hdw}. 
Note that one well-studied class of inhomogeneous field configurations that will relate to inhomogeneous phases in this work are \textit{chiral spirals}, nonlinear waves with expected values of $\sigma$ and $\pi_3$ behaving as a spiral;  i.e., $\left< \sigma+i\pi_3\right> \sim \exp(ik\cdot r)$ \cite{Schon:2000he, Kojo:2009ha}. 
The widespread presence of exotic phases in finite-density models suggests that QCD phase structure could be more complicated than conventionally anticipated, particularly near the QCD phase transition and hypothesized critical endpoint \cite{Nussinov:2024erh}. 
Experimental signatures of moat regimes in heavy ion collisions have been developed, including in Hanbury-Brown-Twiss (HBT) interferometry \cite{Pisarski:2020gkx, Pisarski:2021qof, Rennecke:2023xhc, Fukushima:2023tpv} and enhanced dilepton production \cite{Nussinov:2024erh}. 

In this work, we compute the phase diagram of lattice spin and gauge models with $\zz$ symmetry (the center of the QCD gauge group SU(3)) and $\cCK$-symmetric complex couplings, which mimic a chemical potential and induce a sign problem. Due to $\cCK$ symmetry, we can often construct dualities between complex $\zz$ models and sign problem-free chiral $\zz$ models. The duality is a complex-chiral generalization of the Kramers-Wannier duality of the Ising model. Next, we generalize the real-space renormalization group (RG) for use in systems with chemical potential, and we use it to compute the phase diagram of the complex and chiral $\zz$ models. One class of these models, chiral $\zz$ spin systems, exhibit $\zz$ analogues of chiral spirals: an imaginary analogue of chemical potential induces a phase in which $\zz$ variables exhibit sinusoidal modulation along a given direction. We predict the phase diagrams for broad classes of lattice $\zz$ spin and gauge models with real and imaginary chemical potentials for in three and four dimensions, determining which have $\zz$ spirals. We also explore the extension of our results to $\zn$ models with $N > 3$ and to SU($N$) models with $N\ge 3$.

The outline of this paper is as follows. In the remainder of \sec{intro},  we introduce the $\zz$ models that form the focus of our paper. In \sec{techniques}, we introduce two key techniques used in this paper, Kramers-Wannier duality and the Migdal-Kadanoff renormalization group (RG), and discuss their extensions to complex and chiral models. In \sec{results}, we use the Migdal-Kadanoff RG to compute the phase diagram of our models. In \sec{pt-z3-discussion}, we discuss our results, highlighting how different RG schemes lead to qualitatively different phase diagrams in finite-density models. We also extend our results to certain SU($N$) lattice models. Finally, in \sec{outlook}, we offer concluding remarks and outlook.

\subsection{Complex and chiral $\zz$ models}
We define 1D complex and chiral $\zz$ spin models with actions 
\begin{align}\label{eq:models}
	S_{c}[s_{j}]&=\sum_{j=1}^{M}\left[\frac{J}{2}\left(s_{j}s_{j+1}^{*}+s_{j}^* s_{j+1}\right)+\frac{\theta}{\sqrt{3}}\left(s_{j}s_{j+1}^{*}-s_{j}^* s_{j+1}\right)\right]
	\nonumber\\
	S_{\chi}[s_{j}]&=\sum_{j=1}^{M}\frac{\tilde J}{2}\left(e^{i\tilde\theta}s_{j}s_{j+1}^{*}+e^{-i\tilde\theta}s_{j}^ *s_{j+1}\right)\,,
\end{align}
respectively. Here, $s_j$ are $\zn$ spin variables taking values $\exp[2\pi n/N]$, the parameter $j$ runs over the $M$ lattice sites, and the parameters $(J,\theta)$ and $(\tilde J, \tilde \theta)$ are real-valued. In $S_c$, the complex interaction term $\theta$ induces a nonzero density. The parametrization we use for $S_c$ is not the standard one, which can be obtained from $S_\chi$ via the analytic continuation $\tilde\theta \to -i\mu$. It is easy to see that $\mu \sim \theta$ for small $\mu$, and the general relation between parametrizations is straightforward. However, the $(J,\theta)$ parametrization is more convenient for RG calculations.

To define higher-dimensional complex (chiral) spin models, we impose the interaction with $\theta\neq 0$ ($\tilde\theta\neq 0$) in only a single direction, leaving the other $d-1$ transverse directions as standard-nearest neighbor interactions with couplings $J$ ($\tilde J$). In all dimensions, the complex spin models are a model of $\zn$ particles in the presence of a chemical potential. 

\change{
	To define gauge models, we replace site-based spins with link-based gauge fields, and we replace nearest-neighbor spin interactions with Wilson plaquette interactions:
	\begin{align}\label{eq:gauge-models}
		S_{c}[u_{j}]&=\sum_{p}\left[\frac{J}{2}\left(u_p + u_p^*\right)+\frac{\theta}{\sqrt{3}}\left(u_p-u_p^*\right)\right]
		\nonumber\\
		S_{\chi}[u_{j}]&=\sum_{p}\frac{\tilde J}{2}\left(e^{i\tilde\theta}u_p+e^{-i\tilde\theta}u_p^*\right)\,,
	\end{align}
	where the sum is taken over all plaquettes $p$ in the 2D plane, and the variables $u_p$ are the standard plaquette variables, i.e.  the product of four $\zz$ link variables around a plaquette.}

To define higher-dimensional gauge models, we replace site-based spins with link-based gauge fields, and we replace nearest-neighbor spin interactions with Wilson plaquette interactions. Just as in the spin models, we take the complex (chiral) term $\theta$ ($\tilde\theta$) to be nonzero in only a single preferred plane. In the complex gauge models, the $\theta$ term corresponds to a background electric field that is real in Minkowski space and imaginary in Euclidean space; this field induces a sign problem on the lattice. The $\tilde\theta$ interaction in the chiral gauge models gives rise to a real magnetic field in both Euclidean and Minkowski space. 

\paragraph{Phase structure of real and chiral $\zn$ models.}
The phase structure of standard $\zn$ models (where $\theta$ and $\tilde \theta$ are both zero) has been extensively investigated \cite{Banks:1977cc, PhysRevD.19.3698, PhysRevD.19.3715, PhysRevB.16.1217, Ukawa:1979yv, Cardy:1981fd}. Less is known about general chiral and complex $\zn$ models, though many studies of chiral spin models have been carried out. In three and higher dimensions, chiral $\zz$ spin models have an infinite number of stable inhomogeneous phases that are commensurate with the underlying lattice in the low temperature (small $\tilde J$) region. In a given inhomogeneous phase, $(d-1)$-dimensional sheets of spins, each characterized by a certain expectation value, are layered along the chiral direction \cite{PhysRevB.24.398,Yeomans1981, PhysRevB.24.5180, selke_monte_1982,  Yeomans1982, Bak1982, mccullough_mean-field_1992,Asorey:1993ya}. These phases are $\zz$ analogues of chiral spirals, with each phase corresponding to a particular wave number. Similar spirals occur in $\zn$ with $N\ge 4$. It is likely that the $N\to\infty$ limit is smooth, although to our knowledge this has not been investigated. The phase diagrams of the chiral spin models exhibit a fractal structure called a Devil's flower, by analogy with the well-known Devil's staircase \cite{10.1007/BF02418423}.
The behavior of $\zz$ spin models in 2D is a special case, and is closely tied to the physics of the BKT transition \cite{Berezinsky:1970fr, Kosterlitz:1973xp}. For example, there is a nontrivial critical point on the positive real axis associated with a second-order phase transition, whereas higher dimensions have first-order transitions. It is known that a technique called the Migdal-Kadanoff real-space RG (which we discuss below in detail) does only a fair job of capturing 2D $\zn$ critical behavior \cite{Jose:1977gm}. 

\section{Techniques}\label{sec:techniques}
Next, we introduce two key techniques that we will use in our analysis: Kramers-Wannier duality in \secs{kw-real}{kw-complex}, and the real-space RG in \secs{mkrg-real}{mkrg-complex}. 

\subsection{Kramers-Wannier duality}\label{sec:kw-real}

Kramers-Wannier duality maps the 2D Ising model onto itself, interchanging low- and high-temperature behavior \cite{Kramers:1941kn, Savit:1979ny}. It is simple to establish this duality by mapping the Boltzmann weights of one model to the character expansion of the other \cite{Ukawa:1979yv}.\footnote{The weights of a path integral are simply the exponential of the discretized action $w(\vec{x}) = \exp[-S(\vec{x})]$. A character expansion is simply an expansion of a function on some group $\cO = \sum_i c_i \chi_i$ as a linear combination of the characters $\chi_i$ of the group's irreducible representations.} This is simple to see in the 2D Ising model, defined as
\begin{align}
	\cH = -J \sum_{j,k} \sigma_j \sigma_k\,,
\end{align}
where $J$ and $h$ are couplings, $\{j,k\}$ are nearest-neighbor sites, and $\sigma_i$ are $\bZ_2$ spin variables.  Using shorthand notation to write nearest-neighbor interactions $\sigma_j \sigma_k = \sigma_\ell$, we can write the character expansion of each Ising model Boltzmann weight as
\begin{align}
	w(\sigma_\ell) = e^{K \sigma} = \cosh K + \sigma_\ell \sinh K \,. 
\end{align}
where we use the $\bZ_2$ group characters $\{1,\sigma_\ell\}$ with hyperbolic functions as coefficients. The Boltzmann weights themselves are $\exp(\tilde K\sigma_0) = \{\exp(\tilde K), \exp(-\tilde K)\}$. We only care about the relative weighting of the Boltzmann weights, so we can construct a duality 
\begin{align}
	\exp(-2\tilde K) \leftrightarrow \tanh K\,.
\end{align}
This duality maps the small-$K$ region of the Ising model, which is paramagnetic with zero magnetization, to the large-$K$ region of the Ising model, which is ferromagnetic with nonzero magnetization. The fixed point of the duality transform is the critical point of the Ising model. This duality extends to operators \cite{Kadanoff:1970kz}, and is the most well-known example of order-disorder duality.

This duality extends to other $\bZ_2$ and $\zn$ models generally. The form of the duality is dimension-dependent, as shown in \tab{dualities}. In spin systems, nearest-neighbor spins are connected by links. The Kramers-Wannier dual of a link interaction on a square lattice in two dimensions is a link on the dual lattice perpendicular to the original link. In three dimensions, the dual of a link is a plaquette in 3D, and a cube in 4D. Thus, the dual of a $\zn$ spin system is a $\zn$ spin system in two dimensions, a gauge theory in three dimensions, and a theory of fundamental plaquettes interacting around a cube in four dimensions. Similarly, the dual of a $\zn$ gauge theory is a $\zn$ spin theory in three dimensions and a $\zn$ gauge theory in four dimension. For a concise treatment of Abelian lattice duality, see e.g. \cite{Ukawa:1979yv}. 

\begin{table}\def\arraystretch{1.3}
	\centering
	\begin{tabular}{ |c|c| } 
		\hline
		\rowcolor{PapayaWhip} \null\quad {\bf Dimensionality}\quad\null & {\bf Duality of real $\zn$ models}
		\\ \hline
		2D & Spin systems $\leftrightarrow$ Spin systems
		\\ 
		3D & Spin systems $\leftrightarrow$ Gauge theories
		\\
		4D & \null \quad  Gauge theories $\leftrightarrow$ Gauge theories \quad\null
		\\ \hline
	\end{tabular}
	\caption[Kramers-Wannier duality]{Kramers-Wannier duality for real lattice $\zn$ models.}\label{tab:dualities}
\end{table}

\subsection{Complex-chiral extension of Kramers-Wannier duality}\label{sec:kw-complex}
Building off studies of similar $\cPT$ $\zn$ models in \refscite{Meisinger:2012va, Meisinger:2013zba} and studies of duality in other $\cPT$-symmetric lattice models \cite{Ogilvie:2018fov}, it is straightforward to show that complex and chiral $\zn$ models exhibit an extension of the Kramers-Wannier duality. We begin by examining the character expansion and Boltzmann weights of our chiral and complex models, which allows us to demonstrate a duality between these models. Next, we highlight the regions of parameter space in which the complex models have a sign problem-free dual form.

\paragraph{Character expansion of $\zz$ links.}
\change{The partition functions for the spin and gauge models are defined by sums over $\zz$ spins and links in the usual way, as:
	\begin{align}
		&Z_{\rm spin}=\sum_{\{s_j\}}\exp\left[\sum_{\ell}A(s_\ell)\right]\,,
		&&Z_{\rm gauge}=\sum_{\{u_\ell\}}\exp\left[\sum_{p}A(u_p)\right]\,.
	\end{align}
	where the fundamental variables in the two sums over configurations are the site variables $s_j$ and link variables $u_\ell$ respectively; the total action is a sum over functions $A(s_\ell)$ of the link variables $s_\ell$ in the spin case and the plaquette variables $u_p$ in the gauge case. We can write these
	expressions as a sum over products of Boltzmann weights:
	\begin{align}
		&Z_{\rm spin}=\sum_{\{s_j\}}\prod_{\ell}\exp\left[A(s_\ell)\right]\,,
		&&Z_{\rm gauge}=\sum_{\{u_\ell\}}\prod_{p}\exp\left[A(u_p)\right]\,.
	\end{align}
The character expression for any Boltzmann weight $w(s_\ell)=\exp [A(s_\ell)]$ has the form
}
\begin{equation}\label{eq:link-action}
	w(s_\ell)=a+bs_\ell+cs_\ell^*\,,
\end{equation}
\change{with character coefficients} $a,b,c \in \mathbb C$. 
\change{The character coefficients represent the $\zz$ Fourier transform of the weight functions \cite{Ukawa:1979yv,Kapustin:2014gua}.}

\change{Let us now examine the form of the character coefficients and Boltzmann weights for our models introduced in \sec{intro}. } The symmetries of a link action $A(s_\ell)$ typically are inherited by their corresponding Boltzmann weights $w(s_\ell) = \exp A(s_\ell)$. 
For a conventional $\zn$ model, the link action must be real, implying invariance under complex conjugation $\cK$. We also require invariance under charge conjugation $\cC$, which takes $s_\ell \to s_\ell^*$. These two conditions fix \eq{link-action} to 
\begin{align}
	\change{w_{\rm H}(s_\ell)}=a+b(s_\ell+s_\ell^*)\,,
\end{align}
where $a,b\in \mathbb R$. This Boltzmann weight \change{is positive} for $b>-a/2$. 

To formulate the chiral $\zz$ nearest-neighbor action, we remove the condition that \eq{link-action} is invariant under $\cC$ and only require the action to be real, finding that
\change{
\begin{align}\label{eq:chiral-link-action}
	w_{\chi}(s_l)=1+z s_\ell+z^* s_\ell^*\,,
\end{align}
where we have normalized $a$ to one and parametrized $b$ as $z = x+iy \in \mathbb C$. }

The complex $\zz$ action requires invariance under combined $\cCK$ without imposing invariance under $\cC$ or $\cK$ individually. Under these conditions, \eq{link-action} must satisfy
\change{
\begin{align}\label{eq:complex-link-action}
	w_c(s_\ell^*)=w_c^*(s_\ell)\,,
\end{align}
}
This bears close resemblance to the $\cCK$ symmetry condition in quantum mechanics, $V(x) = V^*(-x)$ \cite{Bender:1998ke}. Under these conditions, $\{a,b,c\} \in \bR$ in \eq{link-action}, but \change{$w_c(s_\ell)$} is complex as the $\zz$ spin variables \change{$s_\ell$} are complex.  

\paragraph{\change{Duality conditions.}} The character expansion of the Boltzmann weights in the \change{complex} $\zz$ model is
\begin{equation}
	w_c(s_\ell)=\exp\left[{J\over 2}\left(s_\ell+s_\ell^*\right)+{\theta\over \sqrt 3}\left(s_\ell-s_\ell^*\right) \change{-J}\right]\,,
\end{equation}
\change{where we have added a constant $-J$ to $A_c$, which normalizes $w(1)=1$. The weights of the complex model are then
	\begin{align}
		&w(1) = 1\,,
		&& w(e^{2\pi i/3}) = e^{-J/2 + i\theta}\,,
		&& w(e^{4\pi i/3}) = e^{-J/2-i\theta}\,.
	\end{align}
The character coefficients are 
\begin{align}\label{eq:complex-weights}
	&a = 1\!+\!2e^{-\frac{3J}{2}}\cos\theta\,,
	&&b = 1\!-\!2e^{-\frac{3J}{2}}\cos\left(\theta-\frac{2\pi}{3}\right)\,,
	&&c = 1\!-\!2e^{-\frac{3J}{2}}\cos\left(\theta-\frac{4\pi}{3}\right)\,.
\end{align}}
\change{We parametrized the character coefficients of the $\zz$ chiral model in \eq{chiral-link-action} as 
\begin{align}\label{eq:tildes}
	&\tilde a = 1\,,
	&&\tilde b = z\,,
	&&\tilde c = z^*\,.
\end{align} 
As in the Ising model, we want to identify the Boltzmann weights with dual representation character coefficients, implying that}
\begin{align}
	z = e^{-J/2 + i\theta}\,. 
\end{align}
\change{This relation is involutive because the $\zz$ Fourier transform is involutive, and it is straightforward to check that the real character coefficients of the complex model are the weights of the chiral model. Note that the normalization condition $\tilde a =1$ for chiral models is equivalent to the normalization condition $a+b+c = 1$ for complex models. 
}

We see that the high-temperature (small-$z$) behavior of the chiral spin model is dual to the low-temperature ($J \gg 1$) expansion of the complex spin model. Likewise, the low-$T$ ($|z|\to1$) chiral behavior is dual to the high-$T$ $(J \to 0)$ complex behavior. This invertible involution extends the duality of conventional $\zz$ models to the complex plane. \change{We summarize several important mappings in \tab{dualities}. }

\begin{table}\def\arraystretch{1.3}
	\centering
	\begin{tabular}{ |c|c|c| } 
		\hline
		\rowcolor{Honeydew}  & {\bf Complex form} & {\bf Dual real form} 
		\\ \hline
		2D & Spin with chemical potential & Spin with chiral interaction 
		\\ 
		3D & Spin with chemical potential & Gauge with background magnetic field
		\\ 
		3D & Gauge with background electric field & Spin with chiral interaction 
		\\
		4D & Gauge with background electric field & Gauge with background magnetic field 
		\\ \hline
	\end{tabular}
	\caption[Complex-chiral duality]{Generalization of the Kramers-Wannier duality introduced in \sec{kw-real} to $\zn$ models with a non-Hermitian transfer matrix.}\label{tab:dualities}
\end{table}

\paragraph{Sign problem-free region.}
Complex $\zn$ models always have a sign problem, but their chiral duals only have a sign problem outside the triangle defined by
\begin{align}\label{eq:triangle}
	1+2x & >  0,\hspace{0.5 in}
	1-x-\sqrt{3}y >  0,\hspace{0.5 in}
	1-x+\sqrt{3}y > 0\,,
\end{align}
\change{where $z = x+iy$ as before.}
While chiral models are not typically defined outside of this triangle, due to the complex-chiral duality, we see that it is natural to consider chiral models on the full range of $(x,y)$ values. 

\subsection{Migdal-Kadanoff renormalization group}\label{sec:mkrg-real}

The Migdal-Kadanoff renormalization group (MKRG) yields qualitative information about the phase structure of lattice systems \cite{Migdal:1975zf, Migdal:1975zg, PhysRevLett.34.1005, Kadanoff1976}. Due to its simplicity and utility, this technique has been extensively studied and applied to many models \cite{RevModPhys.49.267}. Effectively, we apply a Migdal-Kadanoff RG transform $R$ multiple times to a system, and map out the basins of attraction of the model, which tell us the phase structure of a system. The MKRG is approximate on conventional lattices, and exact on hierarchical lattices \cite{derrida_fractal_1983}. Interestingly, MKRG schemes respect Kramers-Wannier dualities \cite{Kadanoff1976}, and thus give us information about both a given $\zn$ model and its dual.

\begin{figure}
	\begin{center}
		\includegraphics[width = 5 in]{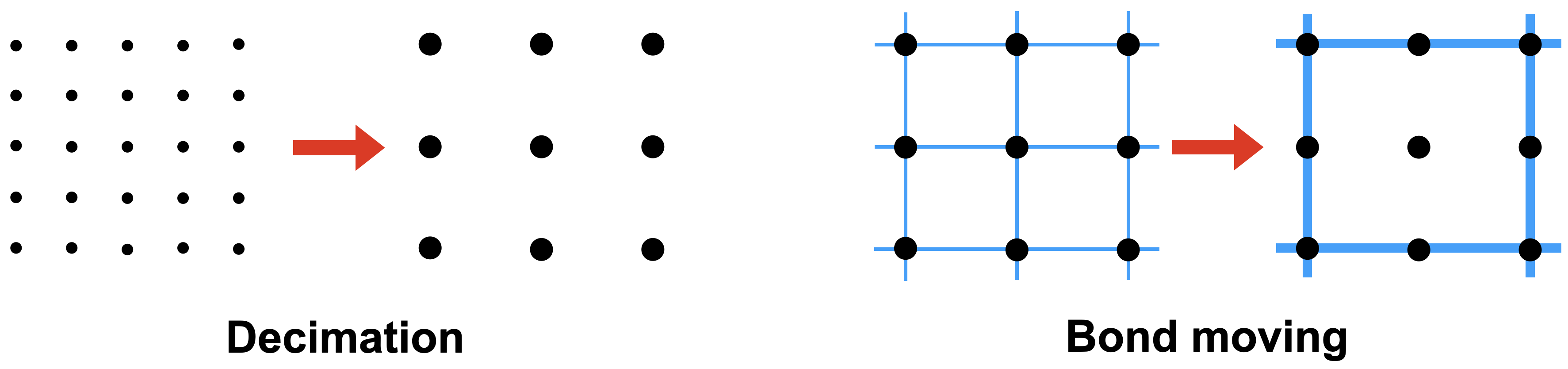}
	\end{center}
	\caption[Migdal-Kadanoff RG transforms]{Schematic representations of Migdal-Kadanoff transforms. In decimation transforms, we eliminate $n^2-1$ out of every $n^2$ lattice sites, as in \eq{decimation}. In bond-moving transforms, we move the entire strength of one bond onto another, as in \eq{bond-moving}.}\label{fig:rg-transforms}
\end{figure}

The MKRG is built from two primary operations: decimation ($D_\lambda$) and bond-moving ($B_\lambda$), as shown in \fig{rg-transforms}. A decimation transformation of a 1D spin lattice changes the lattice spacing from $a$ to $\lambda a$ by integrating out every $(\lambda-1)$ out of $\lambda$ spins, leaving us with lattice points spaced by $\lambda a$ from one another. In a 1D Ising model characterized by nearest-neighbor couplings $J$ and spacing $a$, decimation creates a new Ising model with coupling $J^\prime$ and lattice spacing $\lambda a$. In the 1D Ising model, decimation acts as $\tanh J^\prime = \tanh^\lambda(J)$. 

Decimation can only be carried out analytically in 1D. Bond moving is a technique which changes local interactions in a way which allows decimation to be carried out in higher dimensions. A bond-moving transformation on a spin system moves all the bonds inside a $d$-dimensional hypercube of size $(\lambda a)^d$ to links on the boundaries of the hypercube, changing the strength of an interaction on the boundary of the hypercube from $J$ to $\lambda^{d-1}J$. 

Let us use lowercase $b_\lambda$ and $d_\lambda$ to represent MKRG operations in $\zz$ models. Here, a decimation transform with blocking factor $\lambda = 2$ maps weights $a,b,c$ as:
\begin{align}\label{eq:decimation}
	d_2(a, b, c)= (a^2, b^2, c^2)
\end{align}
and bond moving squares the Boltzmann weights, so
\begin{align}\label{eq:bond-moving}
	b_{2}\left(a,b,c\right) = (a^{2}+2bc,\,c^2+2ab,\,b^2+2ac).
\end{align}
Every MKRG transformation for $\zz$ lattice models can be written as the composition of a sequence of these two operations. 
From these two primitive operations, we can build a large set of MKRG transformations for $\zz$ lattice models. Note that when discussing these transforms, it is convenient to write functional composition as if it were multiplication, so $d_2b_2(a,b,c)$ represents $d_2(b_2(a,b,c))$.

The original Migdal form of the spin model RG has the general form $DB^{d-1}$ \cite{Kadanoff1976}, i.e., bond moving for links in all $d-1$ directions folowed by decimation. This formulation leads to an RG transform
\begin{align}
	J^\prime = R_\lambda(J)=D_\lambda (\lambda^{d-1}J)\,,
\end{align}
Kadanoff \cite{Kadanoff:1970kz} showed bond-moving can also be carried out sequentially along different lattice directions, resulting in an anisotropic system after an RG transformation, e.g.
\begin{align}
	J_k^\prime = R_\lambda(J)=\lambda^{d-k} D_\lambda (\lambda^{k-1}J_k).
\end{align}
for $k=1,\,...\,,\, d$.
This produces a set of MKRG transformations that can have different fixed points for couplings in different directions, but which lead to similar critical behavior for standard lattice systems.  That is, Kadanoff's extension to anisotropic RG flows posits that the RG transformations $B^j D B^{d-1-j}$, with $j=0\dots d-1$, all exhibit the same phase structure for a given spin system. For a gauge theory, Migdal's original RG scheme has $R=D^2 B^{d-2}$, and Kadanoff's scheme extends this to
all permutations of two $D$'s and $d-2$ $B$'s \cite{Kadanoff1976}.

Migdal-Kadanoff RG transforms respect Kramer-Wannier duality \cite{Kadanoff:1970kz}: decimation $D_\lambda$ in a given $\zn$ model is equivalent to bond-moving $\tilde B_{\lambda}$ in its dual model, and likewise for $ B_\lambda$ and $\tilde D_\lambda$. This relation holds because decimation is convolution of Boltzmann weights, which is multiplication of the character expansion coefficients of the weights, and bond-moving is multiplication of weights. The two operations are related by the $\zn$ Fourier transform, which transforms the parameters of one model to the parameters of its dual.

\subsection{Symmetries of the real-space RG for complex and chiral models}\label{sec:mkrg-complex}
\Refcite{PhysRevB.24.5180} pointed out the crucial role of two symmetries in the RG analysis of chiral $\zz$ spin systems, a $\cPT$-type symmetry and a  \textit{Roberge-Weiss symmetry}. These symmetries extend to all $\zz$ chiral and lattice models. To see this, let us construct the Migdal-Kadanoff RG operations for our models, choosing a blocking factor $\lambda = 2$ in \eqs{decimation}{bond-moving}. For the chiral model with weights in \eq{chiral-link-action}, we have bond-moving and decimation transforms
\begin{align}\label{eq:BD-defined}
	&\tilde B(z) =  \frac{2z+z^{*2} }{ 1+2zz^*}\,,
	&&\tilde D(z) = z^2\,,
\end{align}
which impose a normalization $a=1$ \change{and we recall from \sec{kw-complex} that $z= x+iy$}. It is straightforward to show that the chiral transforms $(\tilde D, \tilde B)$ are dual under a $\zz$ Fourier transform to the corresponding complex transforms $(B,D)$, which we can obtain from \eqs{decimation}{bond-moving}. Despite their equivalence, these transforms have different algebraic forms and provide useful cross-checks on one another. 

Roberge-Weiss symmetry is invariance of the partition function under $z\rightarrow \omega z$, where $\omega\in \zz$. This symmetry manifests itself directly in the RG recursion relation. From the explicit forms of $(\tilde D, \tilde B)$ in \eq{BD-defined} we see that
\begin{align}
	\tilde B(\omega z)&=\omega \tilde B(z)\,,\qquad\qquad
	\tilde D(\omega z)=\omega^2 \tilde D(z).
\end{align}
for any $\omega \in \zz$. 
As a consequence, RG transforms $R$ built from a sequence of $\tilde B$'s and $\tilde D$'s obey
\begin{equation}\label{eq:2-20}
	R(\omega z)=\omega^p R(z)\,,
\end{equation}
where $p=0,1,2$. A special case of \eq{2-20} was first found in \refcite{PhysRevB.24.5180} using an argument based on space-dependent $\zz$ transformations of the spins and the behavior of the two-point function. The RG transforms also respect $\cCK$ symmetry:
\begin{equation}\label{eq:pt-in-mkrg}
	R(z^*)=R^*(z),
\end{equation}
which follows from $\tilde B(z^*)=\tilde B^*(z)$ and $\tilde D(z^*)=\tilde D^*(z)$. This relation was first found in \refcite{PhysRevB.24.5180} for chiral spin models. 

\section{Results}\label{sec:results}
We now apply the Migdal-Kadanoff RG and duality to find the phase diagram of our $\zz$ models in \eq{models}. First, it is useful to examine some general features of the fixed points $R(z_0) = z_0$ of $\zz$ models under RG transforms $R$. In dimensions $d\geq 2$, $\zz$ models have a nontrivial fixed point $z_0=x_0+i y_0$ for some $x_0 \in [0,1]$ and $y_0=0$. For standard $\zz$ models, this fixed point separates a large-$J$ region $x<x_0$ from a small-$J$ region $x>x_0$. For chiral and complex $\zz$ models, these phases extend into the complex plane. From \eq{2-20}, we have that
\begin{equation}
	R(\omega z_0)=\omega^p R(z_0).
\end{equation}
In the case $p=0$, we see that $\omega z_0$ is mapped to $z_0$. For $p=1$, we have two additional nontrivial fixed points $\{\omega_0z_0,\omega_0^2 z_0\}$, for $\omega_0=\exp(2\pi i/3)$. For $p=2$, we have $R(\omega_0 z_0)=\omega^2 z_0$ and vice versa, meaning that $\{\omega_0 z_0,\omega_0^2 z_0\}$ form a two-cycle. These features are a natural result of $\cCK$ and Roberge-Weiss symemtries. To summarize:
\begin{align}
	\begin{tabular}{c|c|c|c}
		p & $R(z_0)$ & $R(\omega_0 z_0)$ & $R(\omega_0^2 z_0)$ \\ \hline
		0 & $z_0$ & $z_0$ & $z_0$  \\
		1 & $z_0$ & $\omega_0 z_0$ & $\omega_0^2 z_0$ \\
		2& $z_0$ & $\omega_0^2 z_0$ & $\omega_0 z_0$\\ 
	\end{tabular}
\end{align}
The chiral models also have a high-$T$ fixed point at $z=0$, which is dual to a low-$T$ fixed point at $J=\infty$ with arbitrary $\theta$ in the corresponding complex models. In all cases, lower and upper half-planes have mirror-image phase structure due to \eq{pt-in-mkrg}.

\begin{figure}
	\begin{center}
		\includegraphics[width = 2.0 in]{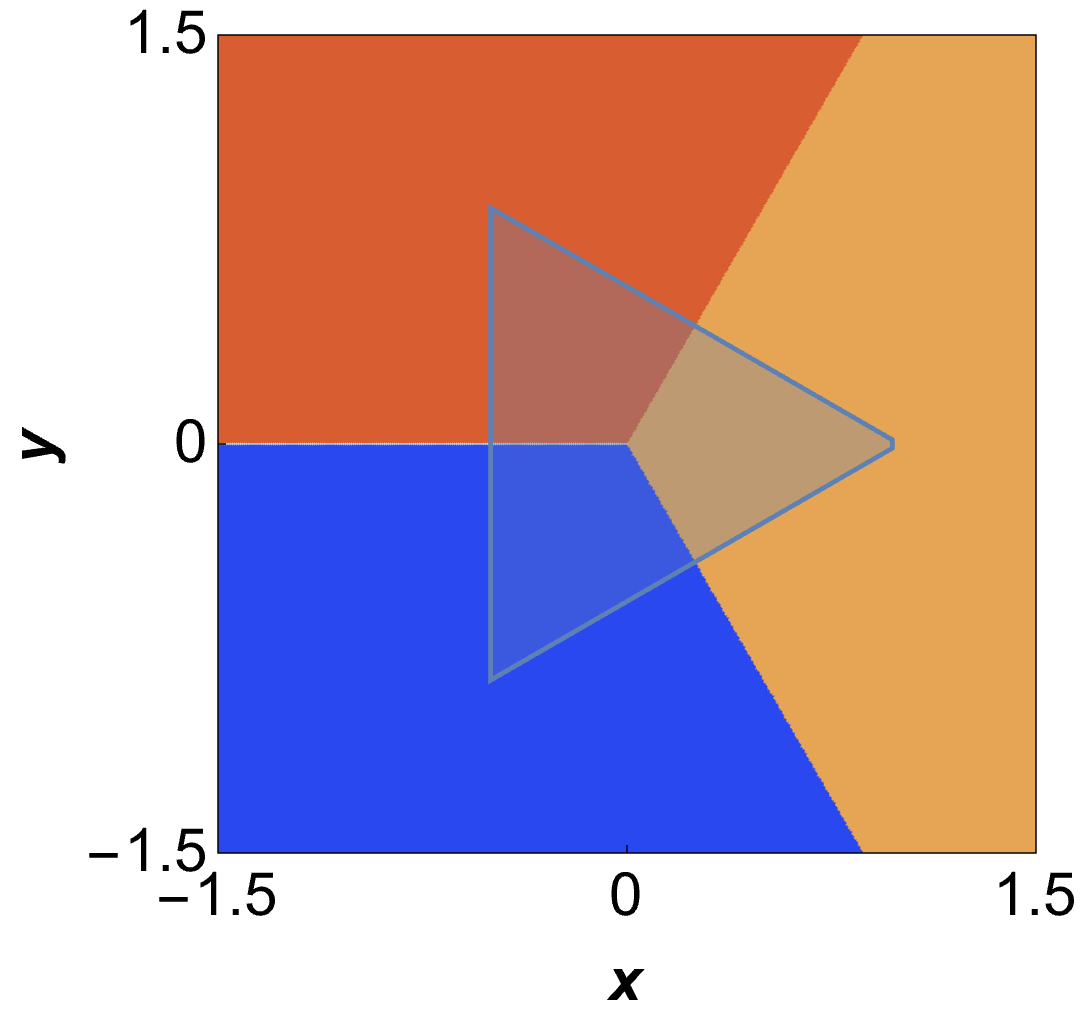}\qquad 
		\includegraphics[width = 2.0 in]{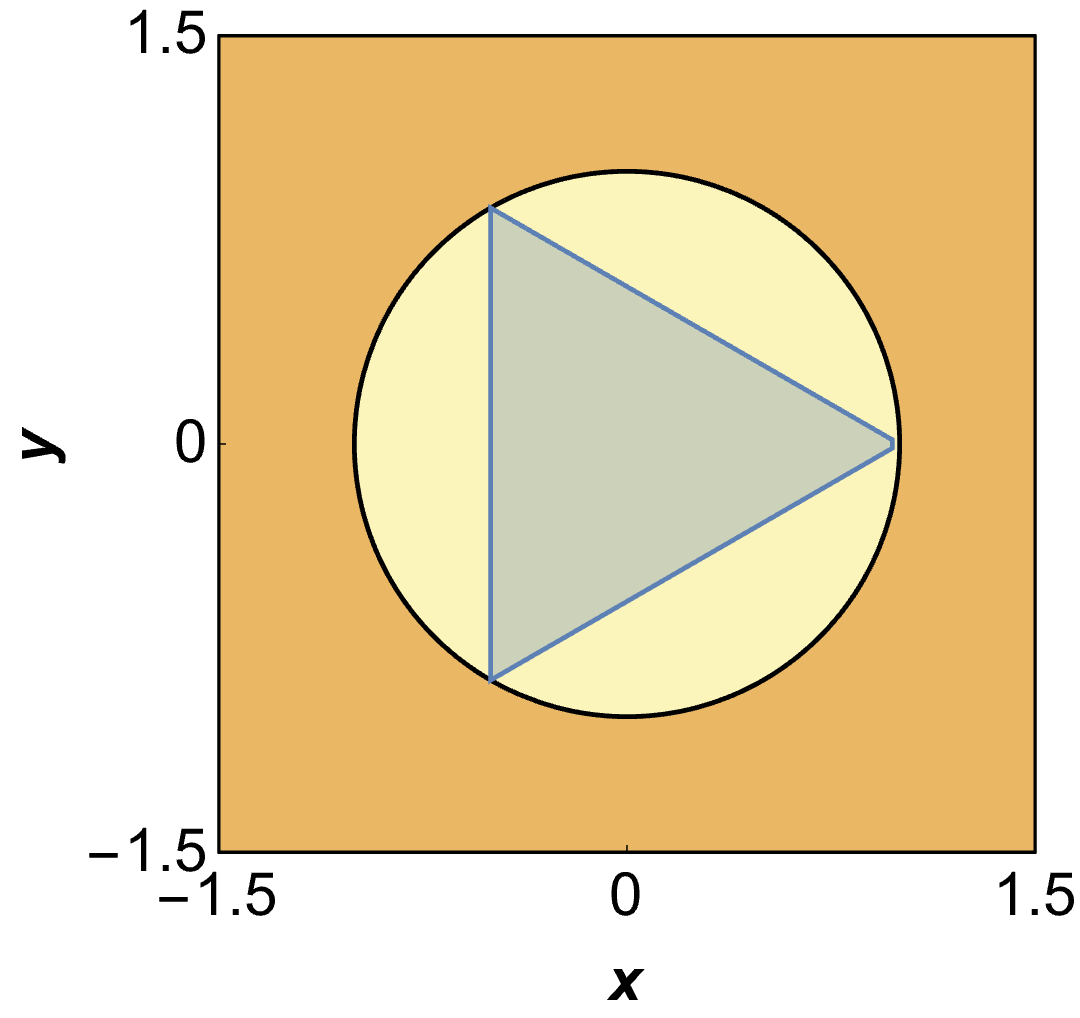}
	\end{center}
	\caption[Phase diagram of 1D complex and chiral $\zz$ spin models]{{\bf 1D complex and chiral spin models.} (Left) 1D complex model with phase structure in the $z=x+iy$ plane obtained from repeated decimation transformations $D \cong \tilde B$. Solid colors correspond to different phases. The light blue triangle in the center represents the region where the action is real.  (right) 1D chiral spin model, obtained from repeated action of $\tilde D \cong B$. All points inside the unit circle eventually map to $z=0$. }\label{fig:1d-models}
\end{figure}

\subsection{1D spin systems}
We determine the phase structure of our models by iterating a given RG transform $R$ on each $z$ a number of times $n$, until the value of $R^n(z)$ is stable for all values of $z$. Typically, we find that $n=6$ is sufficient to give an accurate depiction of the basins of attraction of $R$ and the phase diagram of the model. 

We can express the partition functions of 1D spin models and 2D gauge models exactly in terms of their character weights. For 1D complex $\zz$ spin models, this is
\begin{equation}
	Z=a^N+b^N+c^N
\end{equation}
where $N$ is the number of spins and $\{a,b,c\}$ are the coefficients of the character expansion of the Boltzmann weight given in \eq{complex-weights}; they are also the eigenvalues of the transfer matrix. From the form of $\{a,b,c\}$ given in \eq{complex-weights}, we see that two of the eigenvalues cross when $\theta=\{\pi/3, \pi, 5\pi/3\}$, leading to first-order phase transitions at those values of $\theta$. We confirm this behavior by carrying out an RG analysis with $R(z) = D(z)$, as shown in the left panel of \fig{1d-models}.  The points $z=\{1, \omega_0, \omega_0^2\}$ are all stable high-temperature fixed points, and $z=0$ is an unstable low-temperature fixed point of the complex spin model. The critical lines separating the three phases are the boundaries of the basins of attraction for the high-temperature fixed points, which emanate outward as rays from the $z=0$ along $\theta=\{\pi/3, \pi, 5\pi/3\}$.

We carry out a similar analysis for the 1D chiral spin model, which has partition function 
\begin{equation}
	Z=\tilde a^N+\tilde b^N+ \tilde c^N=1+z^N+z^{*N}
\end{equation}
where $\{\tilde a, \tilde b, \tilde c\}$ are given by \eq{tildes} and $N$ is the number of lattice sites. We display the phase structure obtained from repeated application of the RG transform $R(z)=\tilde D(z)=z^2$ in the right panel of \fig{1d-models}. For $\left|z\right|<1$, the $\tilde a =1$ term dominates the partition function. All values $\left|z\right|<1$ are attracted to the stable fixed point at $z=0$, which is a high-temperature fixed point. Values of $z$ with $\left|z\right|>1$ move off to infinity. RG flows to the origin, to infinity, or along the unit circle are chaotic maps. We can easily see this for $\left|z\right|=1$, where $R$ effectively takes $\theta\to 2\theta \mod 2\pi$ (mapping the unit circle to itself). Taking $x = \sin^2\theta$, we see that this is a particular case of the logistic map $x\to \kappa x(1-x)$ with $\kappa = 4$, which is known to be chaotic. 

In all figures in this section, we draw the sign problem-free region, where the character coefficients $\{a,b,c\}$ are all positive, as a light blue equilaterial triangle given by \eq{triangle}. The vertices of the equilateral triangle are stable fixed points, but other than that, the region of positive weights does not appear to be of fundamental importance to phase structure. Indeed, even in these simple 1D models, we see that restricting the parameter $z$ to the region of positivity in chiral models misses significant aspects of RG flow. 

\subsection{2D complex gauge model}
The phase diagram of the complex gauge model in 2D is similar to that of the 1D complex spin model. The RG flow is generated by $R=D^2=z^4$, so that $R(\omega z)=\omega \tilde B^2(z)$. Thus we again have three stable fixed points  $z=\{1,\omega,\omega^2\}$, where $\omega=\exp(2\pi i/3)$, as in the 1D complex spin model. The phase diagram of the 2D chiral gauge model is effectively the same as that of the 1D chiral spin model: $R=\{z^2,z^4\}$ have similar properties, including the chaotic map on the unit circle; the small change in exponent does not affect RG flow significantly.

\subsection{2D spin systems}

2D chiral and complex spin systems are dual to one another. In Migdal's original formulation of the real-space RG, the natural 2D spin RG transforms are $R=DB$ for the complex model and $R=\tilde D\tilde B=BD$ for the chiral model. Both $DB$ and $BD$ occur naturally in Kadanoff's anisotropic reformulation. In a standard $\zz$ spin model where $z$ is restricted to a real number $x$, these two transformations have a similar phase structure, with some changes in the location of the nontrivial fixed point. 

However, in the chiral and complex models, it is clear from \fig{2d-spin} that the phase diagrams generated by $BD$ and $DB$ differ significantly, producing four and seven phases, respectively. Neither phase diagram shows a Devil's flower, consistent with the results in \refcite{PhysRevB.24.5180}. The fixed points are nonuniversal; there is a nontrivial fixed point $z_0$ along the positive $x$ axis between zero and one at the tip of the orange region, whose value is different for the $BD$ and $DB$. The points $(\omega z_0,\omega^2 z_0)$ form a two-cycle in both cases. 

\subsection{3D spin and gauge systems}

\begin{figure}
	\begin{center}
		\includegraphics[width = 1.75 in]{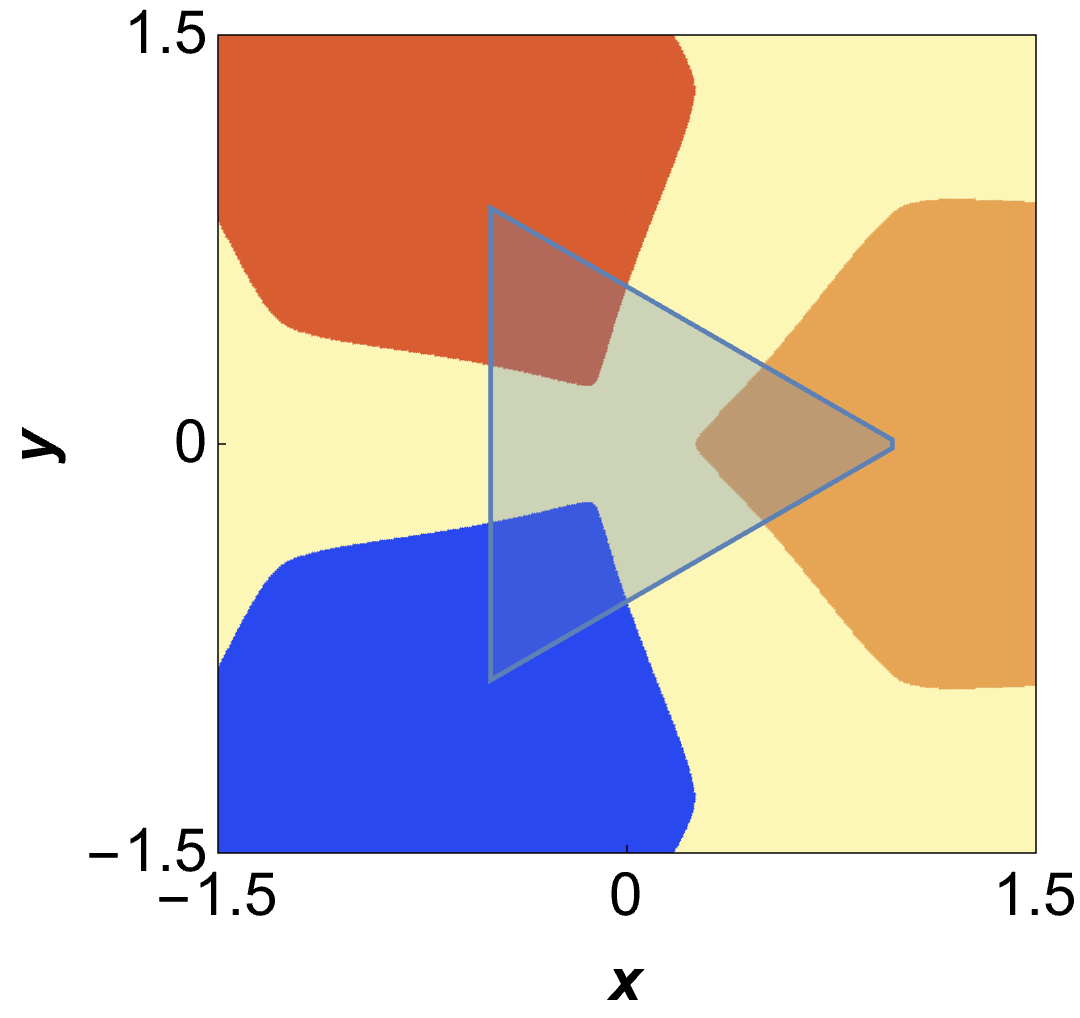}
		\includegraphics[width = 1.75 in]{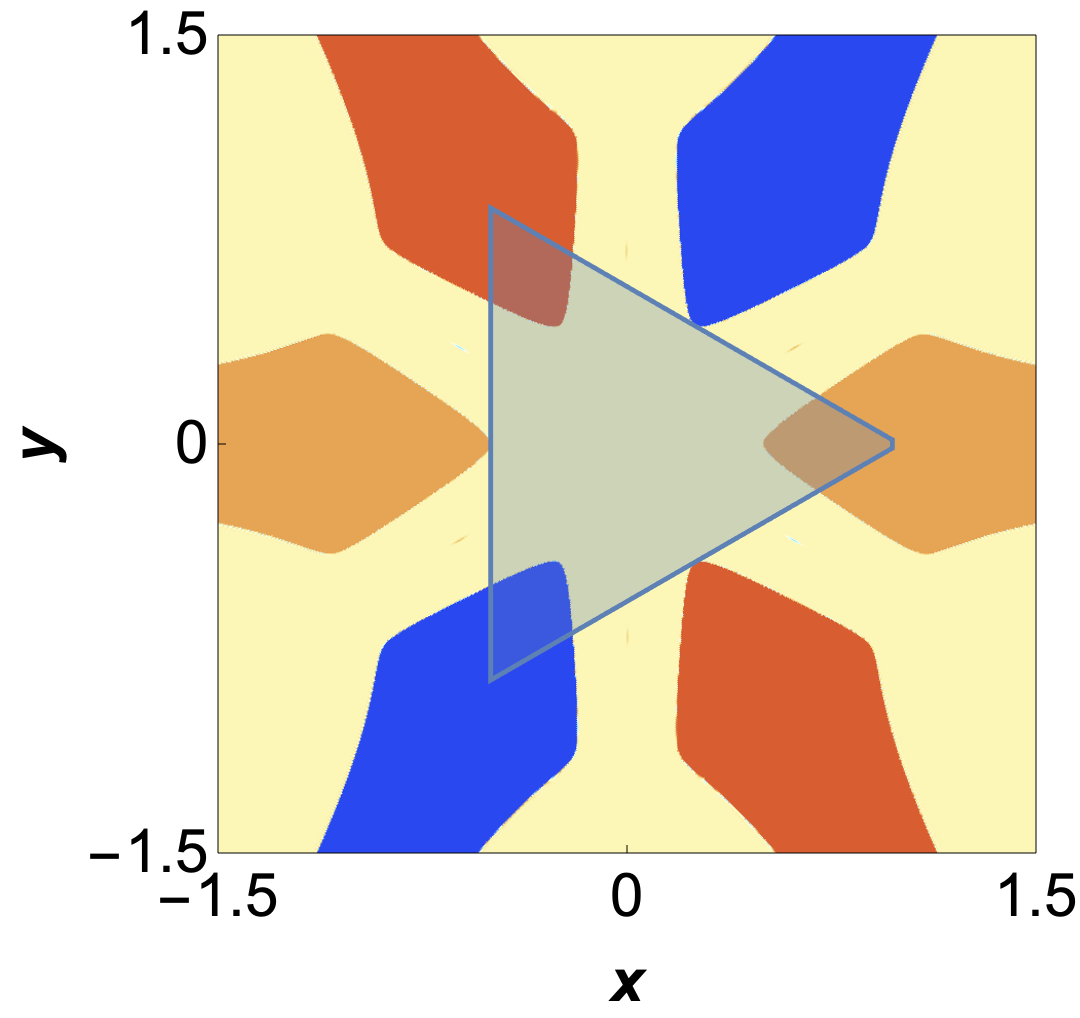}
	\end{center}
	\caption[Phase diagram of 2D complex $\zz$ spin model]{{\bf 2D complex spin model,} analyzed using transforms $BD$ (left) and $DB$ (right).}\label{fig:2d-spin}
	\begin{center}
		\includegraphics[width = 1.75 in]{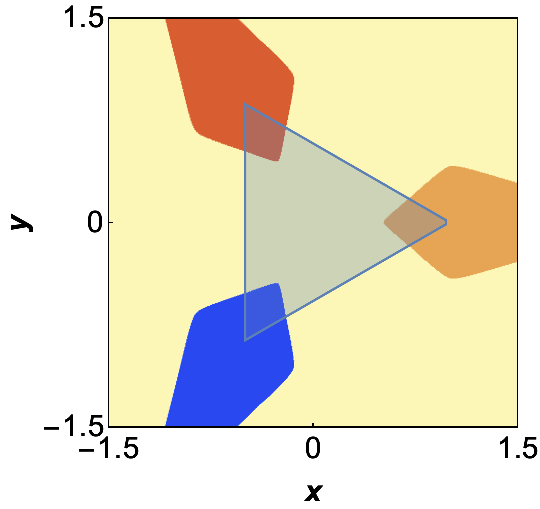}
		\includegraphics[width = 1.75 in]{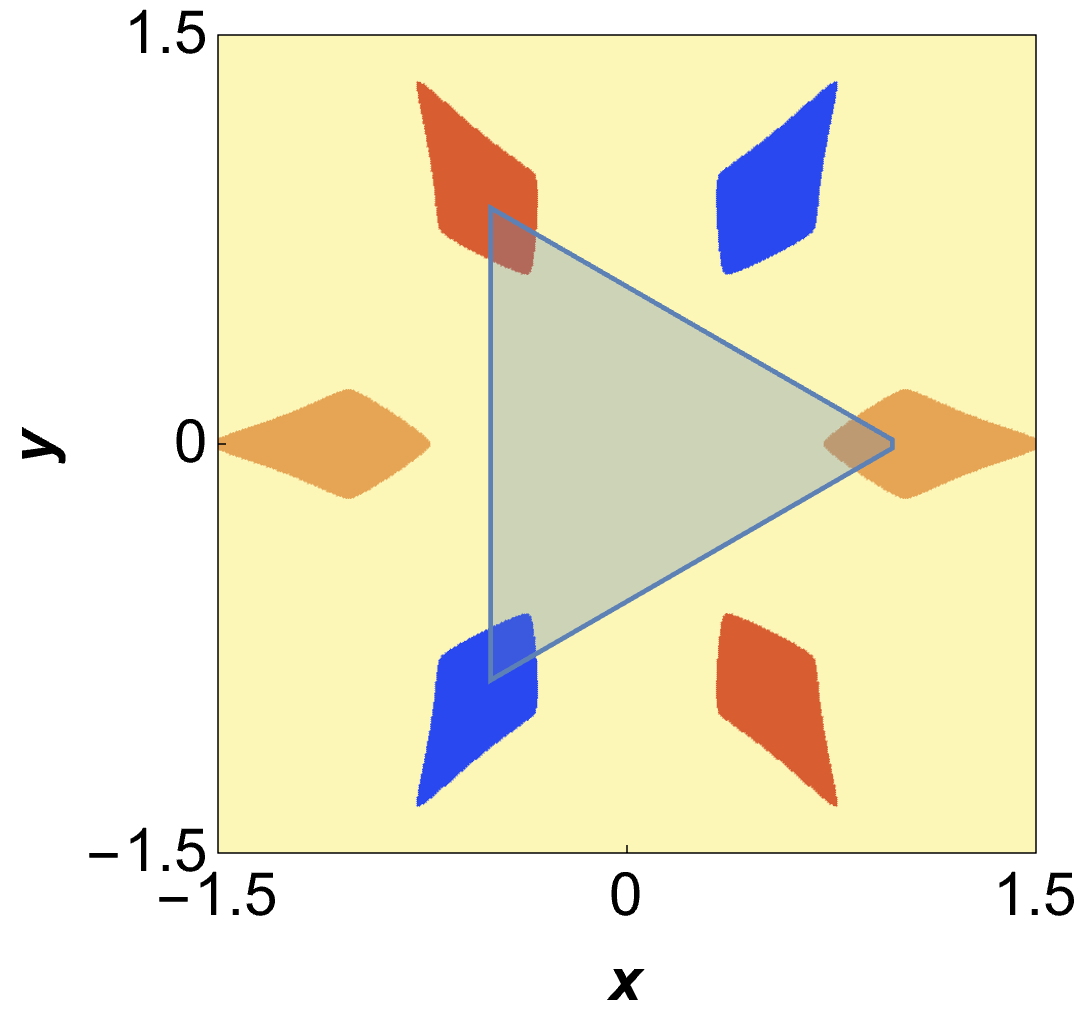}
		\includegraphics[width = 1.75 in]{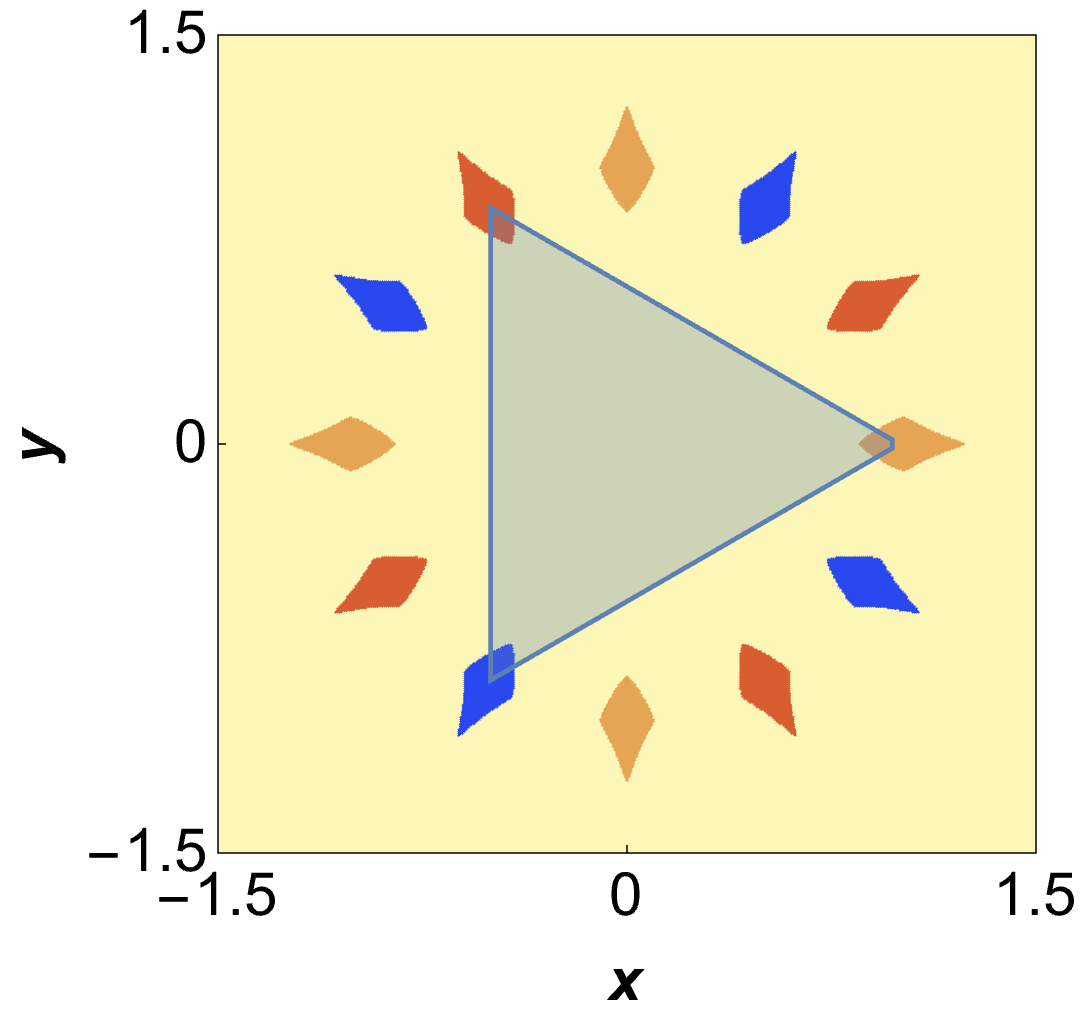}
	\end{center}
	\caption[Phase diagram of 3D complex $\zz$ spin model]{{\bf 3D complex spin models,} analyzed using the transforms (left to right): $BBD$, $BDB$, $DBB$. The phase diagrams are identical to those for 3D gauge models using $\tilde D\tilde D\tilde B$, $\tilde D\tilde B\tilde D$ and $\tilde B\tilde D \tilde D$, respectively.}\label{fig:3d-spin}
	\begin{center}
		\includegraphics[width = 1.75 in]{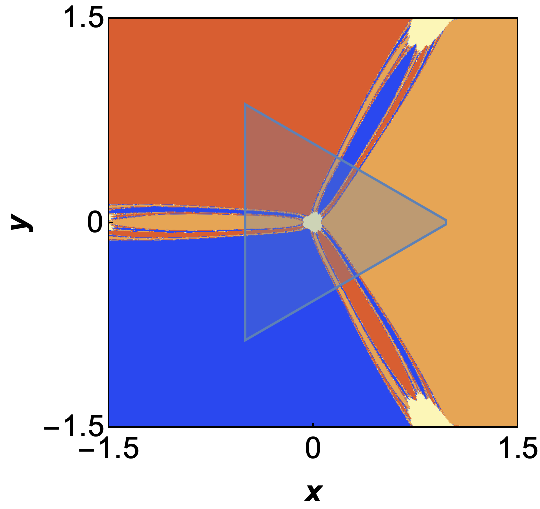}
		\includegraphics[width = 1.75 in]{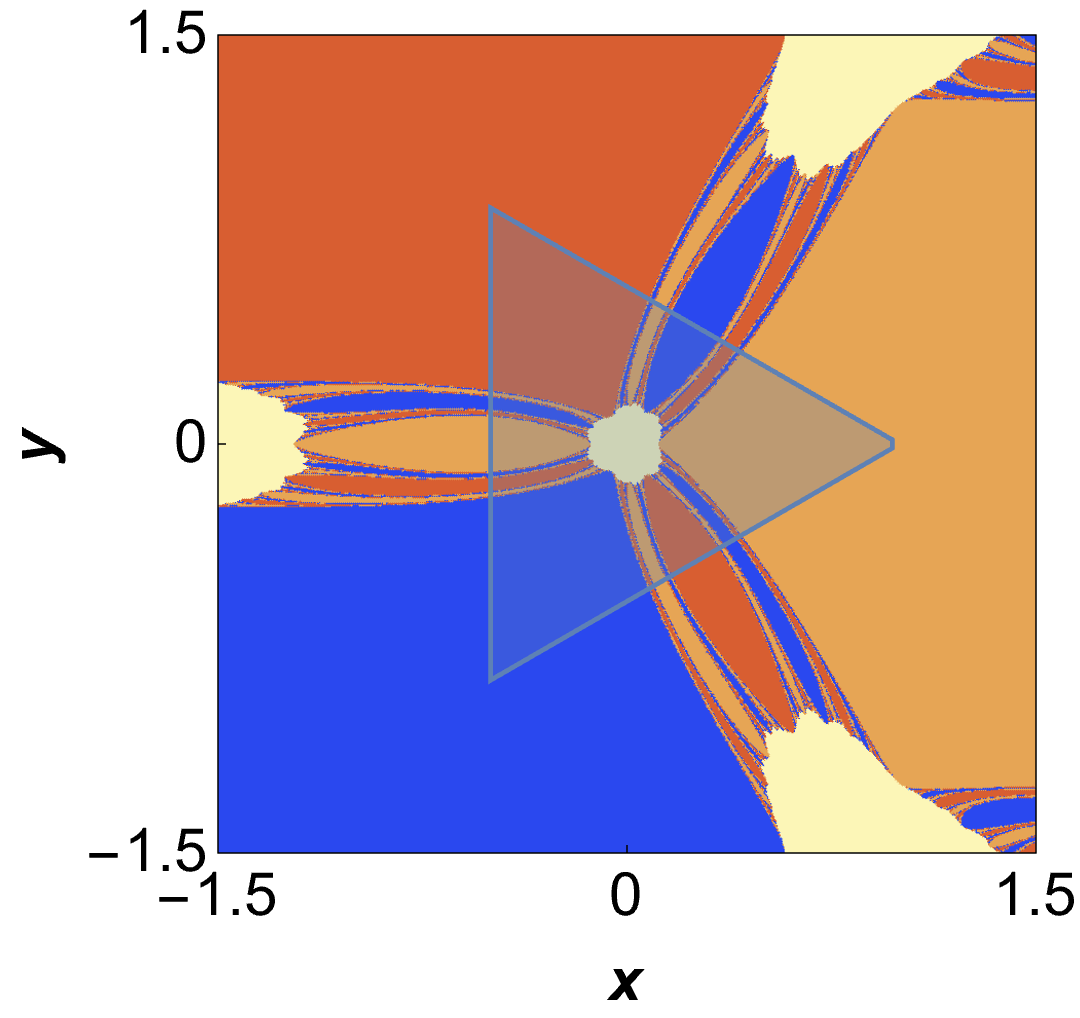}
		\includegraphics[width = 1.75 in]{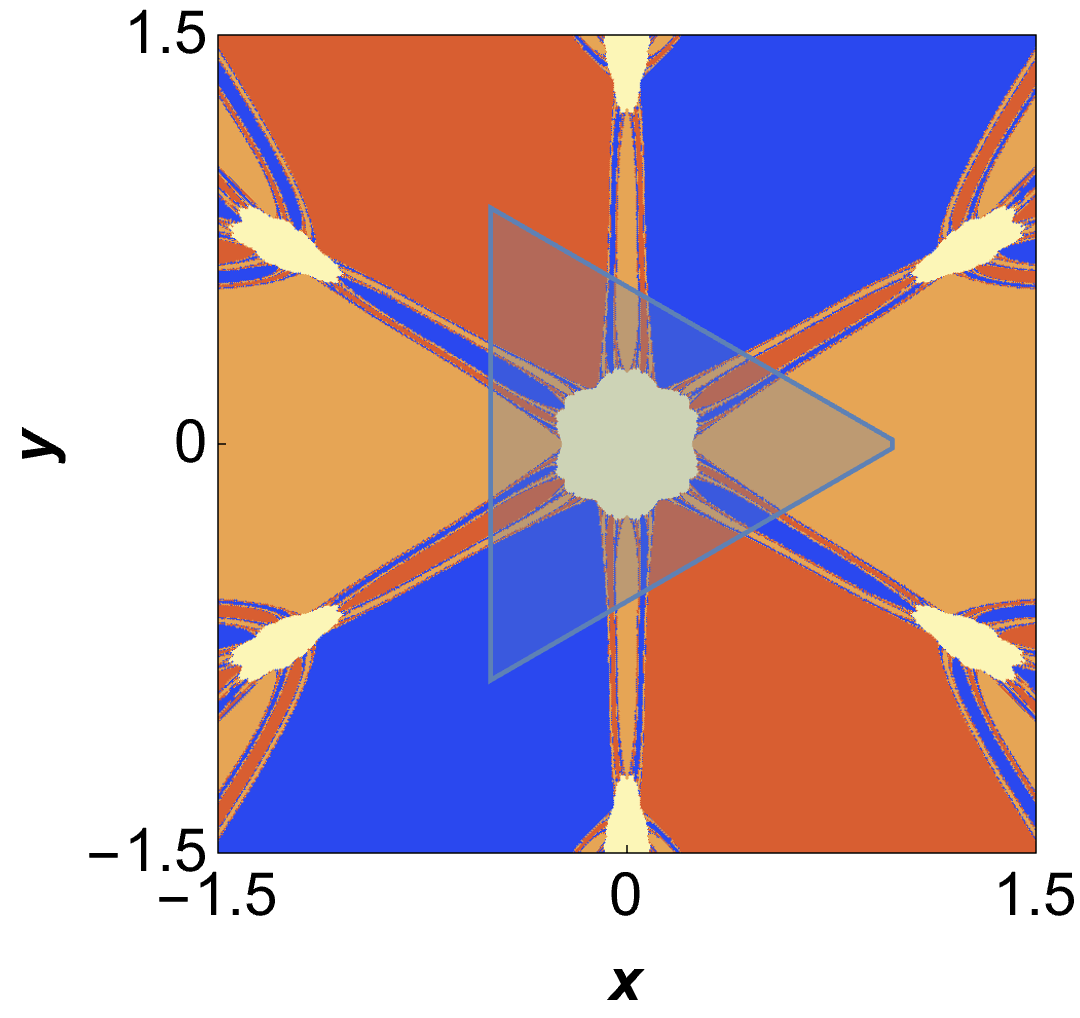}
	\end{center}
	\caption[Phase diagram of 3D complex $\zz$ gauge model]{{\bf 3D complex gauge models,} analyzed using the transforms (left to right): $BDD$, $DBD$, $DDB$. The presence of the Devil's flower phase structure is clear in all three cases. These are identical to the phase diagrams for the 3D chiral spin model.}\label{fig:3d-gauge}
\end{figure}

In \fig{3d-spin}, we plot the phase diagrams of the 3D complex spin model, which is dual to the 3D chiral gauge model. As expected, the location of the nontrivial fixed point on the positive real axis is not universal. As in the 2D complex spin model, the number of phases that appear depends on the order of the basic RG operations, with $BBD$ producing the smallest number of phases (4), and $DBB$ producing the largest number of phases (13). 

In \fig{3d-gauge}, we plot the 3D complex gauge model, which is dual to the 3D chiral spin model. These models exhibit a Devil's flower for each of the transformations $BDD$, $DBD$, and $DDB$. Intriguingly, $BDD$ and $DBD$ have a three-fold symmetry, while $DDB$ has a six-fold symmetry. From the point of view of the chiral spin model, there are an infinite number of commensurate inhomogeneous phases in the low-temperature region, which corresponds to the strong coupling region near $J=0$ for complex models. 

\subsection{4D spin and gauge systems}

\begin{figure}
	\begin{center}
		\includegraphics[width = 1.75 in]{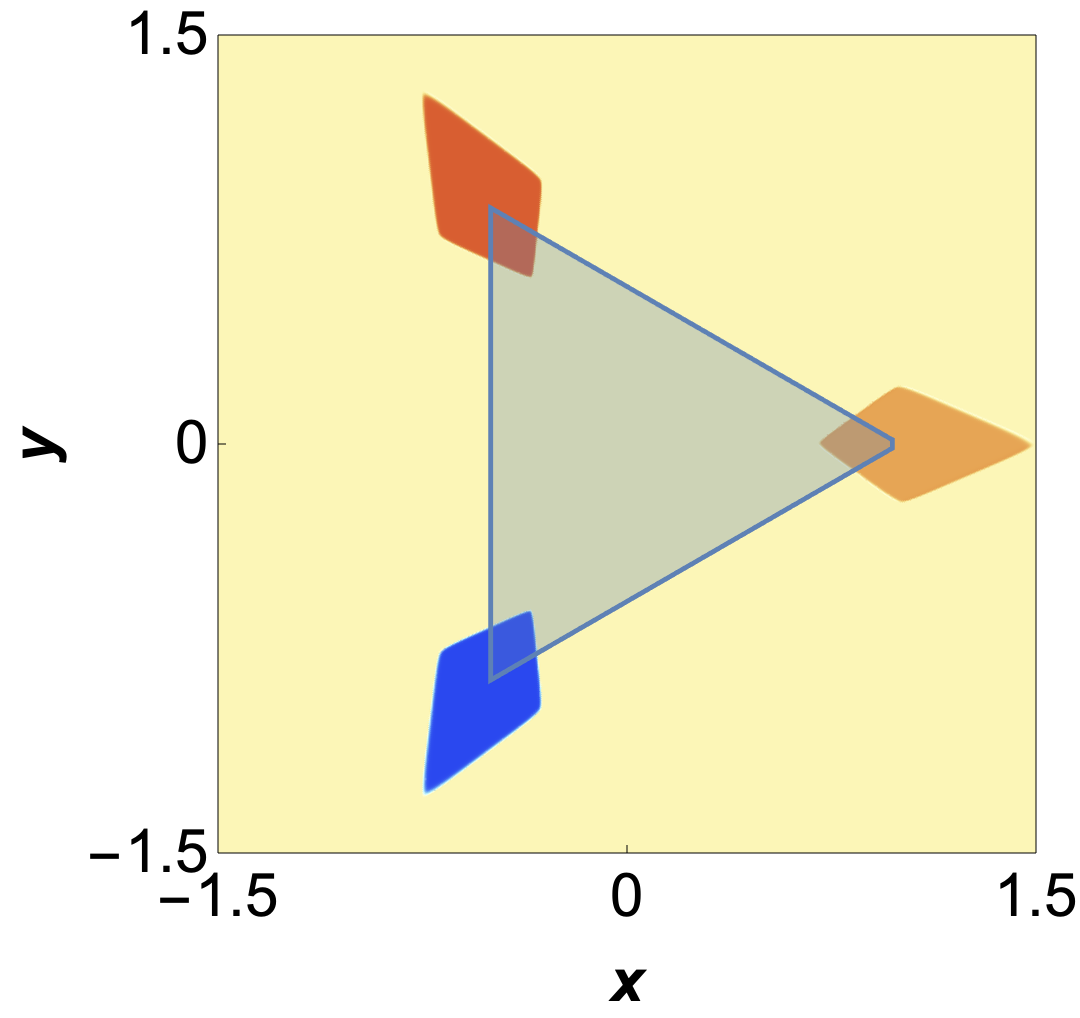}\quad
		\includegraphics[width = 1.75 in]{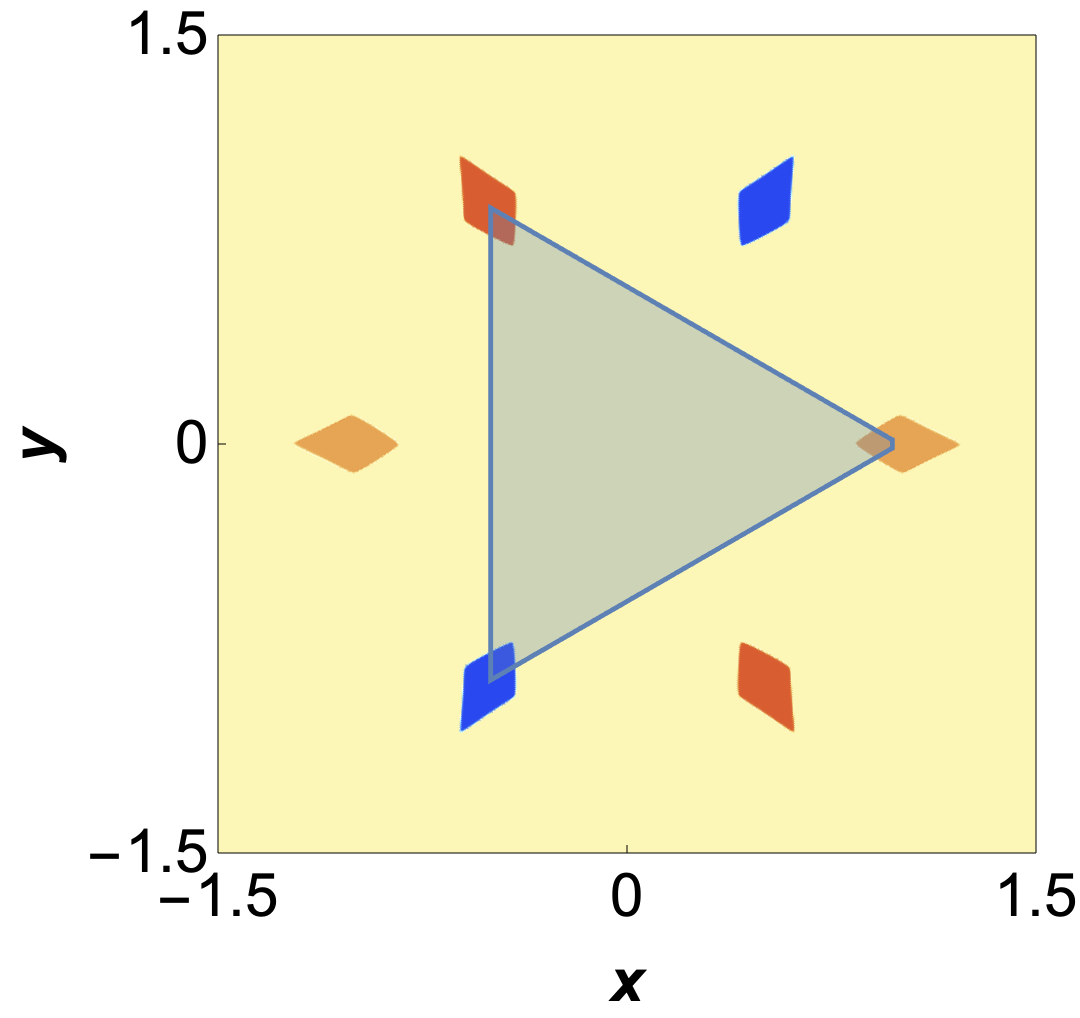}\\
		\includegraphics[width = 1.75 in]{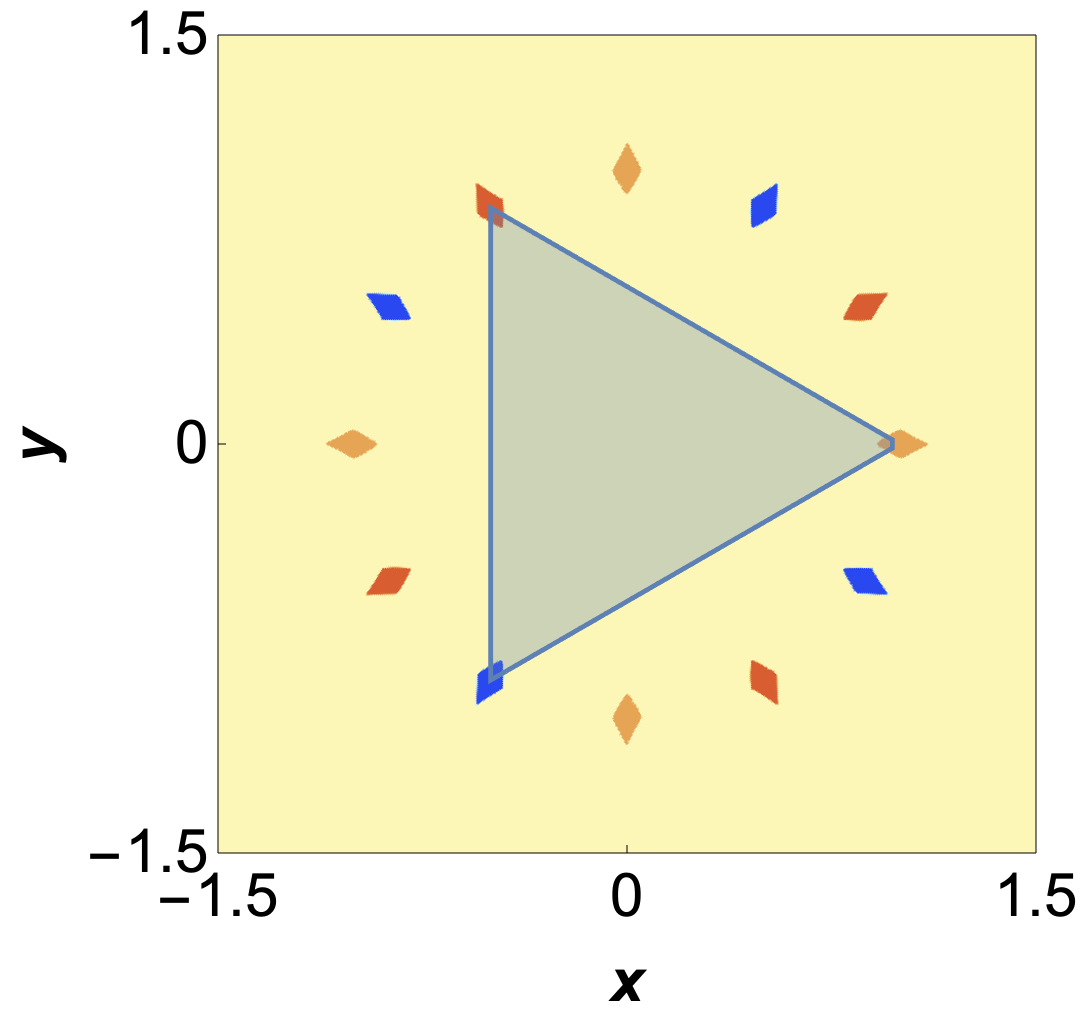}\quad
		\includegraphics[width = 1.75 in]{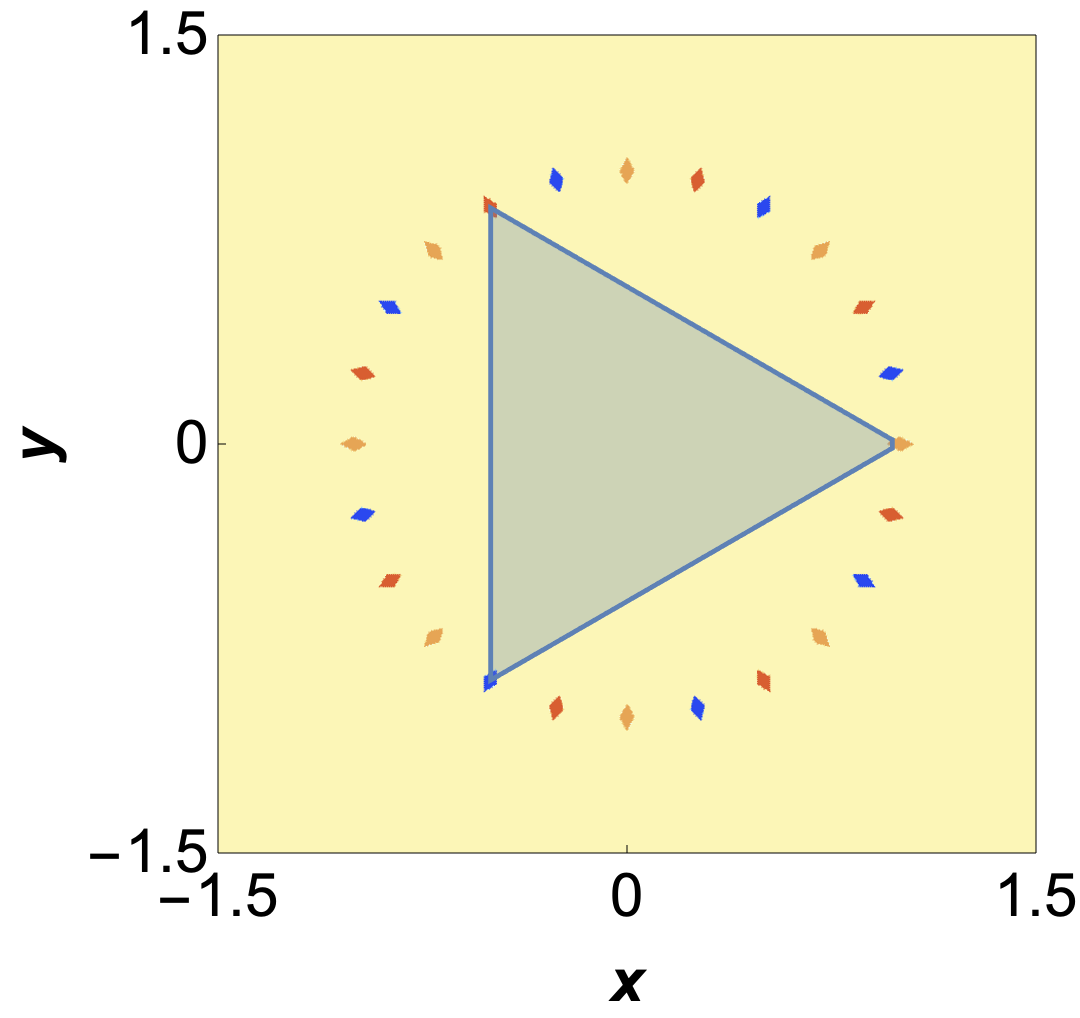}
	\end{center}
	\caption[Phase diagram of 4D complex $\zz$ spin model]{{\bf 4D complex spin models,} analyzed using the transforms (left to right): $BBBD$, $BBDB$ (top row); and $BDBB$, $DBBB$ (bottom row). In the dual chiral model, the fundamental variables are plaquettes interacting around cubes.} \label{fig:4d-complex-spin}
	\begin{center}
		\includegraphics[width = 1.75 in]{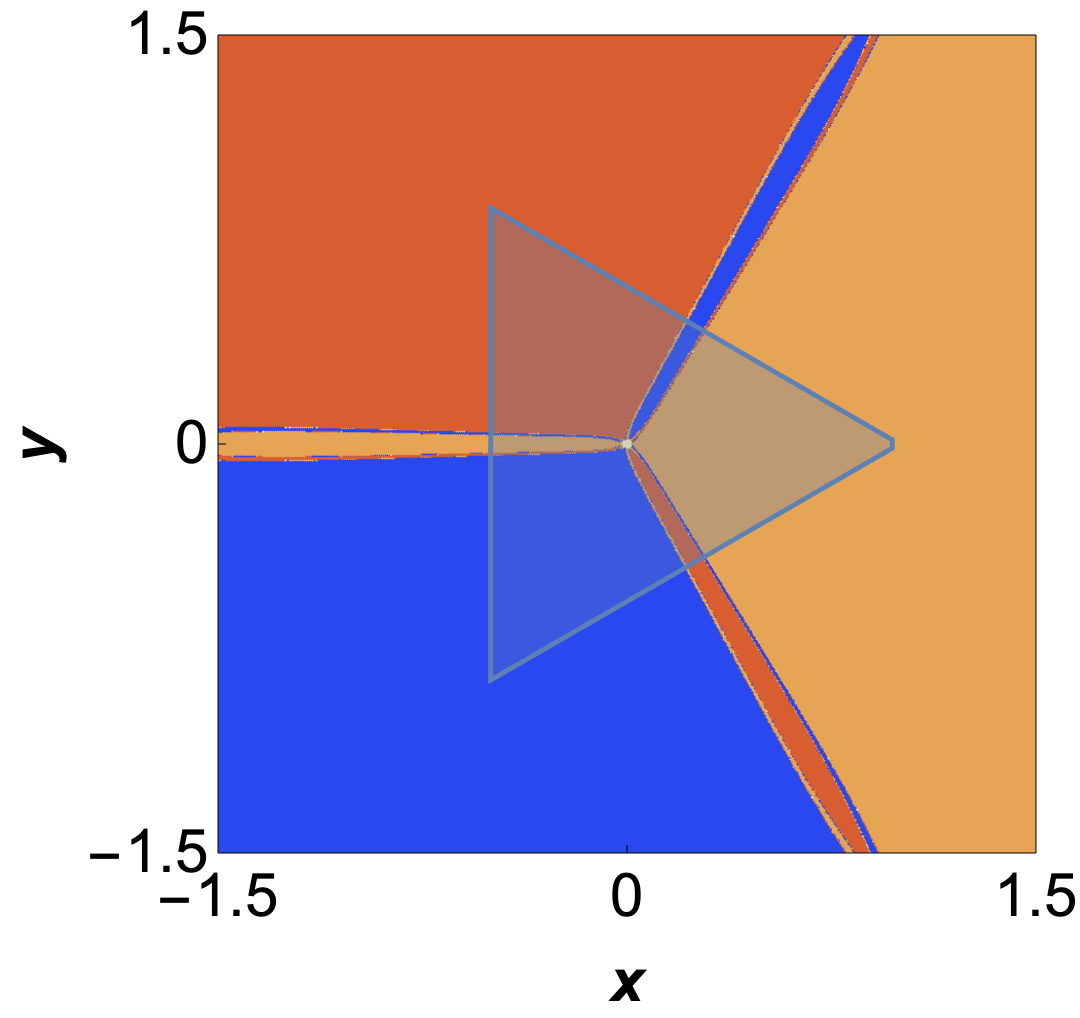}\quad
		\includegraphics[width = 1.75 in]{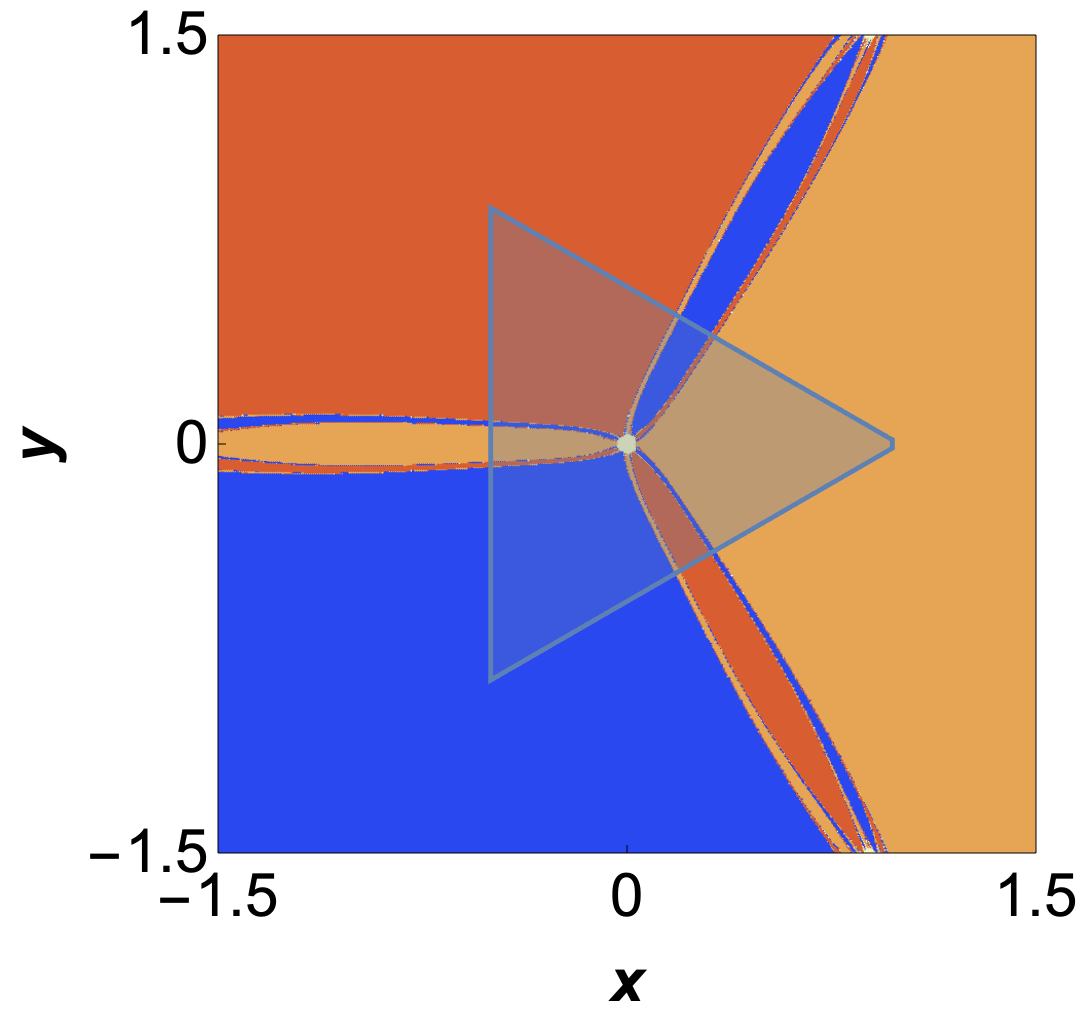}\\
		\includegraphics[width = 1.75 in]{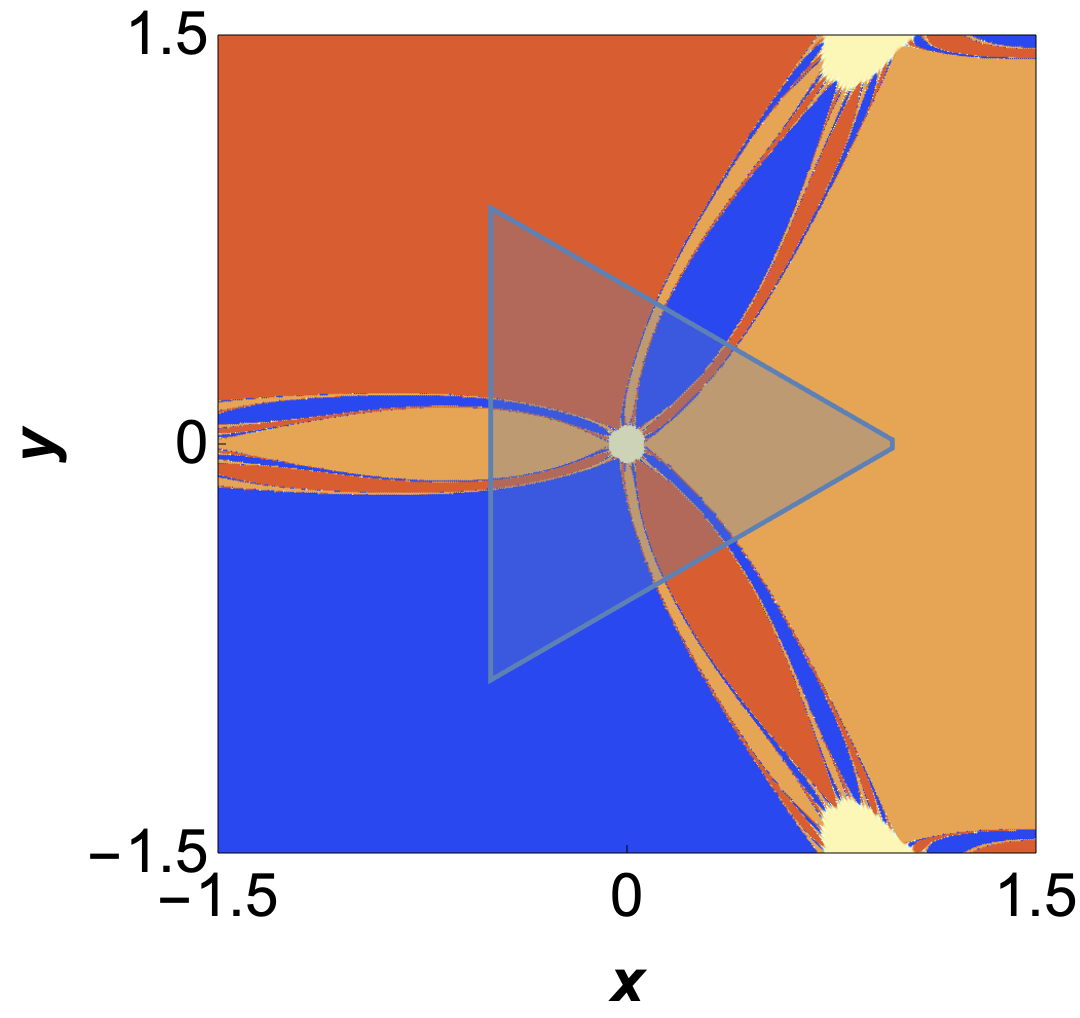}\quad
		\includegraphics[width = 1.75 in]{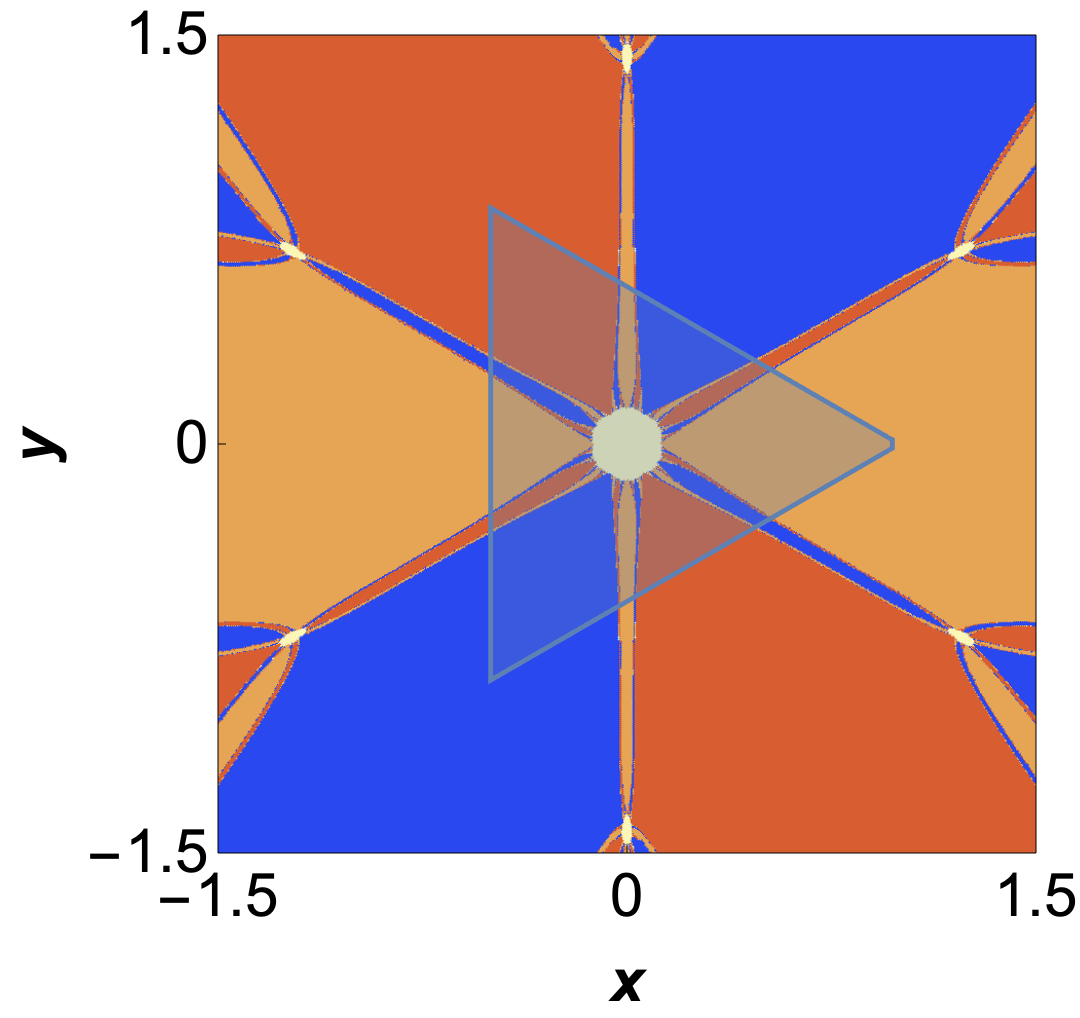} 
	\end{center}          
	\caption[Phase diagram of 4D chiral $\zz$ spin model]{{\bf 4D chiral spin models,} analyzed using the transforms $BDDD$, $DBDD$ (top); and $DDBD$, $DDDB$ (bottom). In the dual complex model, the fundamental variables are plaquettes interacting around cubes.}\label{fig:4d-chiral-spin}
\end{figure}

In \fig{4d-complex-spin}, we see that just as in the 3D models, the order of bond-moving and decimation transforms produces different results for 4D complex spin systems. The RG transform $D B^3$ generates only four regions (one homogeneous and three inhomogeneous). As more $B$ operators are placed to the right of $D$, more phases of smaller sizes appear, with $DBBB$ having 25 different phases, 21 of which lie outside of the triangle of positive weights.

\Fig{4d-chiral-spin} shows the behavior of the 4D chiral spin model. The RG transformation $B D^3$ (equivalent to $\tilde D \tilde B^3$) is the orignal Migdal transformation for this model. The phase diagram takes the form of a Devil's flower, with the critical point on the positive real axis close to the origin. Other permutations of $BD^3$ show similar behavior, albeit with critical points further from the origin. Note that the phase diagram of the transform $D^3B$ has a six-fold symmetry, while the other transforms have a three-fold symmetry.

\begin{figure}
	\begin{center}
		\includegraphics[width = 1.75 in]{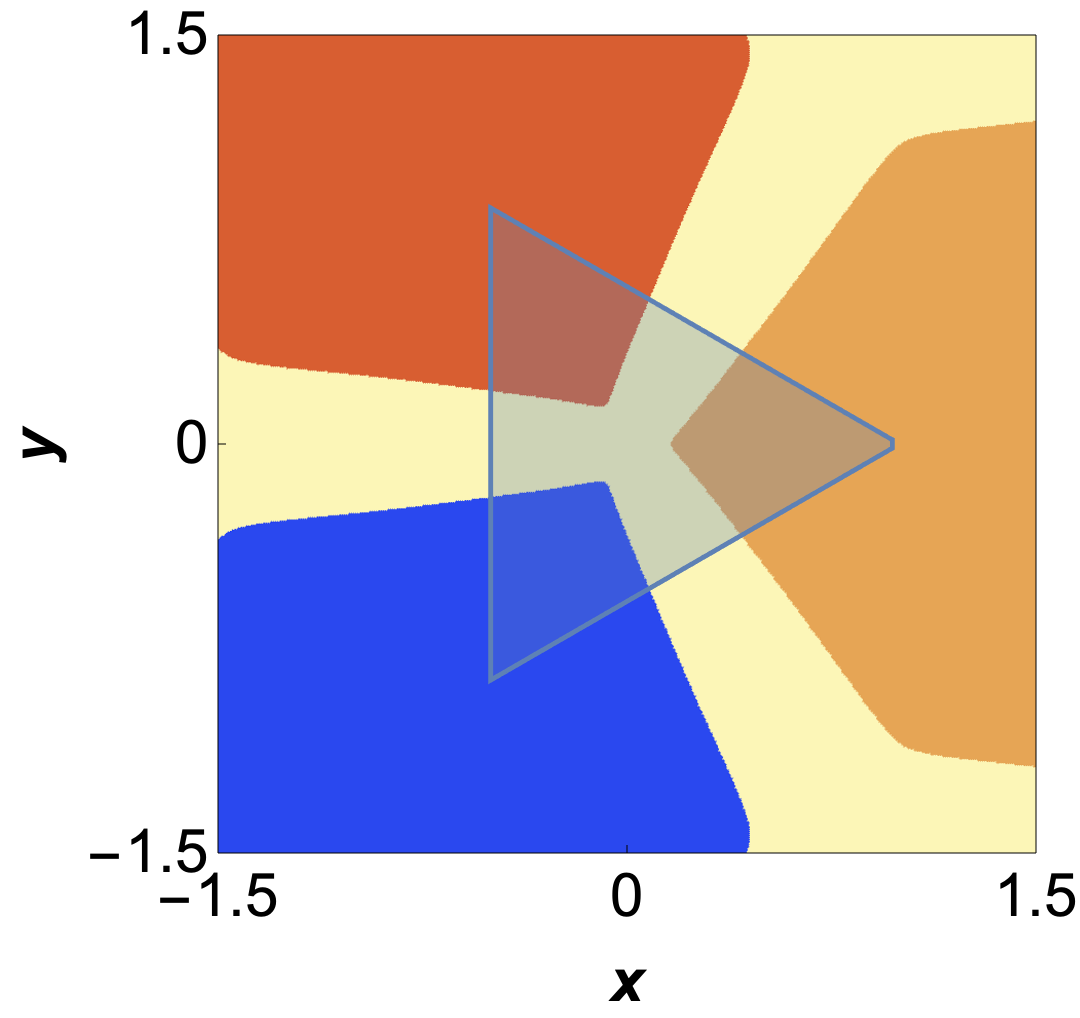}\quad
		\includegraphics[width = 1.75 in]{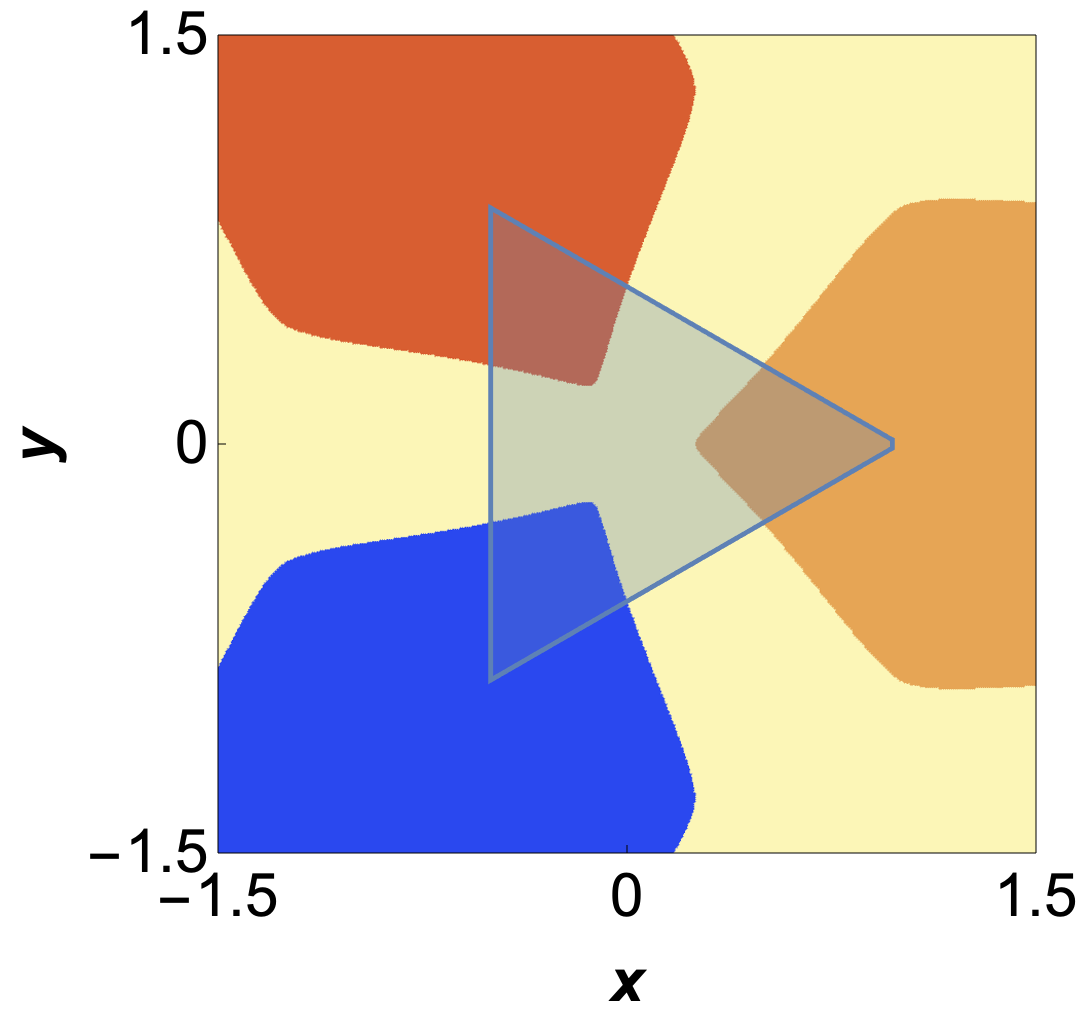}\quad
		\includegraphics[width = 1.75 in]{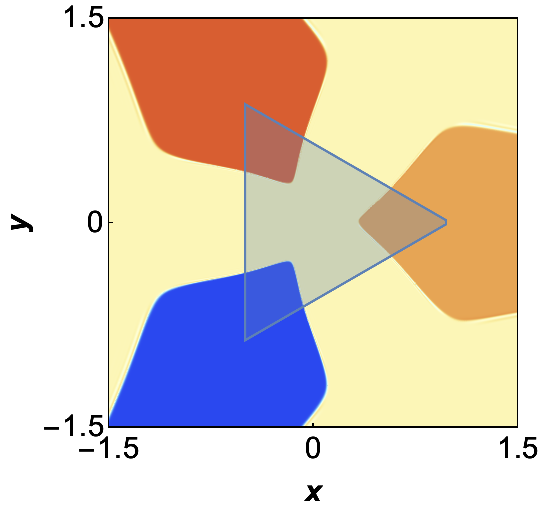}\\
		\includegraphics[width = 1.75 in]{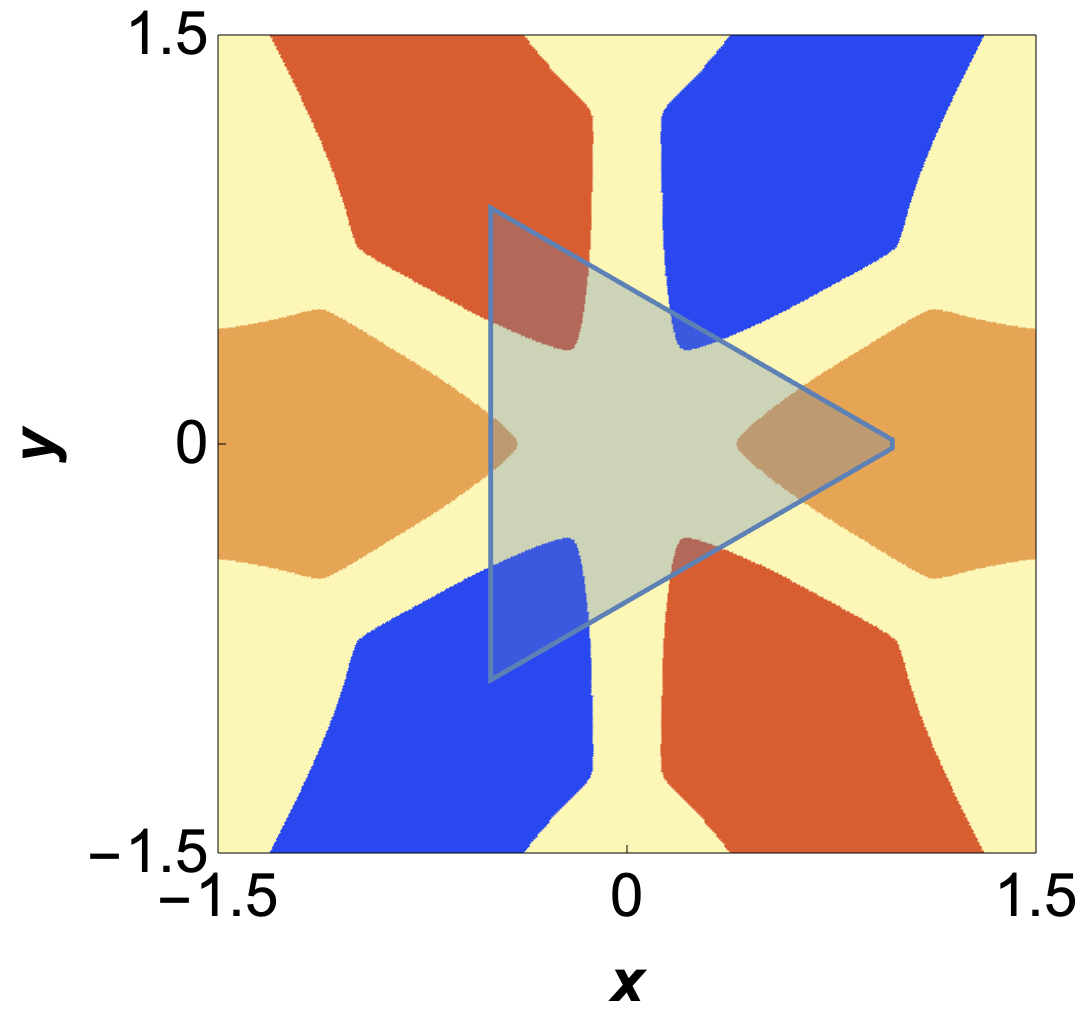}\quad
		\includegraphics[width = 1.75 in]{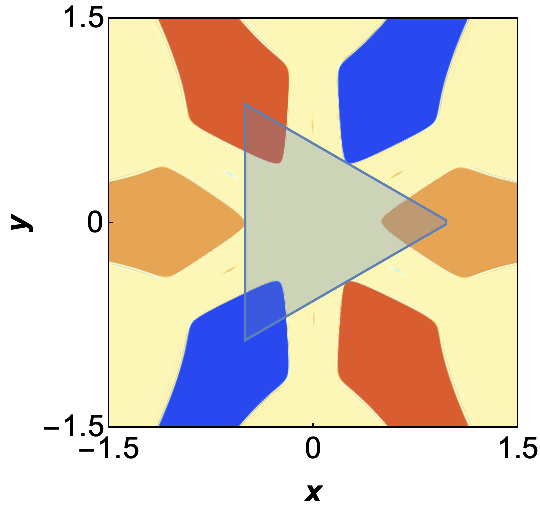}\quad
		\includegraphics[width = 1.75 in]{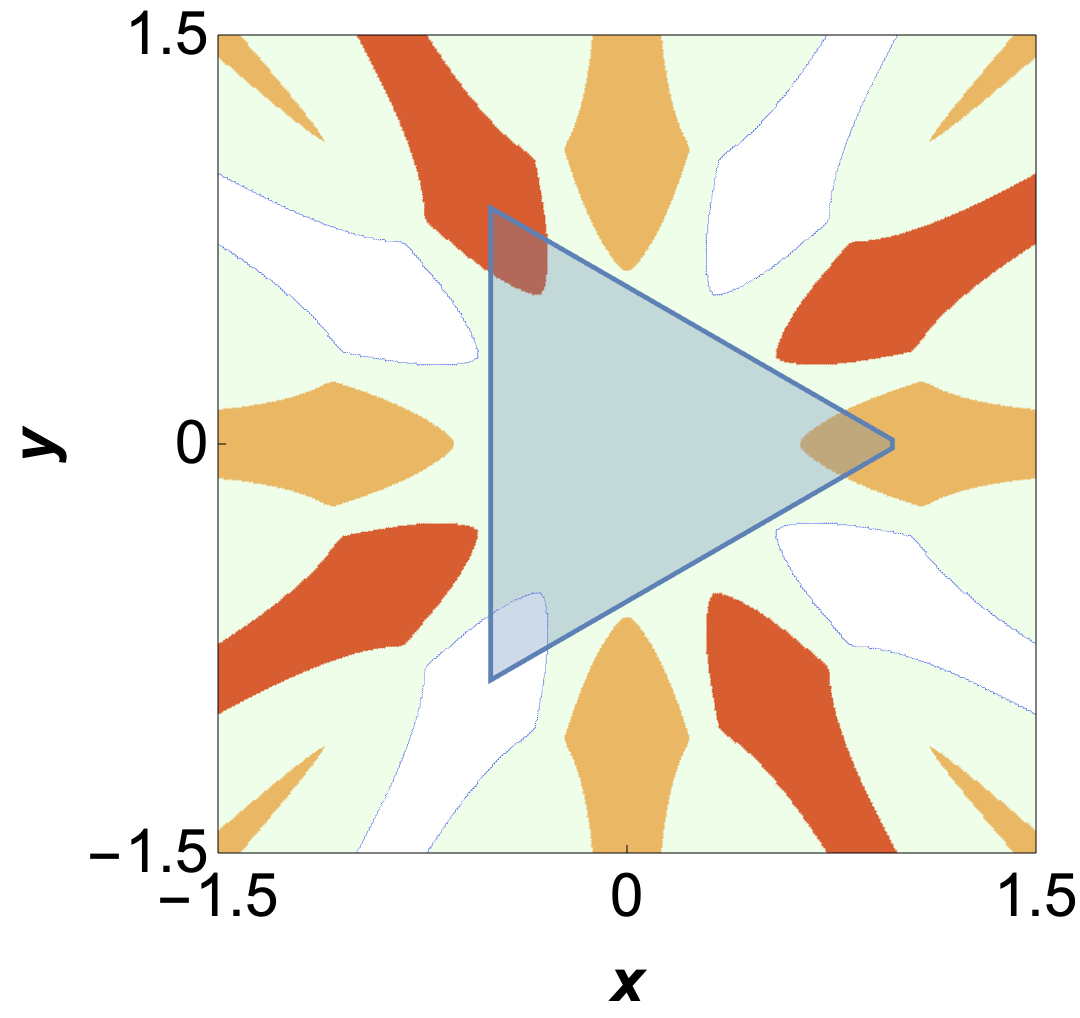}	
	\end{center}
	\caption[Phase diagram of 4D complex and chiral $\zz$ gauge models]{{\bf 4D gauge models,} using the transforms: BBDD, BDBD, and DBBD (1st row); and BDDB, DBDB and DDBB (2nd row).}\label{fig:4d-complex-gauge}
\end{figure}

In \fig{4d-complex-gauge}, we plot the phase diagram of the 4D $\zz$ gauge theory. Using the original Migdal transformation $\tilde D^2 \tilde B^2 $, we see four phases. On the other hand, the transformation $\tilde B^2\tilde D^2$ generates many more phases. 

\section{Discussion}\label{sec:pt-z3-discussion}

Many features of the phase diagrams in \sec{results} are consistent with our intuition from real-space RG analyses of conventional models. All of the spin models have at least four phases for $d\geq 2$, and all of the gauge models have at least four phases for $d\geq 3$, which is the minimum number of phases consistent with three nontrivial fixed points. The location of these nontrivial fixed points is dependent on the form of the RG transforms (i.e. the order of bond-moving and decimation operations), consistent with Kadanoff's interpretation of the real-space RG. However, the phase structures predicted by different RG transforms are far from identical. Here, we discuss a number of their important features.

\subsection{Failure of universality and \change{spurious symmetries}}

The symmetry of phase diagrams often change from threefold, to sixfold, to even higher-fold as we change the order of bond-moving and decimation operations; see e.g. \figs{2d-spin}{3d-spin}. This behavior, to our knowledge, has not been observed in conventional lattice models. \change{Critical behavior usually depends on the symmetries and dimensionality of a system, and our results are} at odds with Kadanoff's interpretation of different orderings as belonging to the same universality class \cite{Kadanoff1976}. Although all Migdal-Kadanoff RG schemes give exact results for appropriate hierarchical models \cite{derrida_fractal_1983}, the phase diagram's symmetry on a normal cubic lattice cannot depend on  the precise scheme used: there is a single correct answer. This difference in symmetry represents a failure of the predictive power of the \change{MKRG to respect the universality within a given model in non-Hermitian systems.}  

The number of phases appears to be tied to the number of $D\equiv\tilde B$ operators appearing to the right of the rightmost $B\equiv \tilde D$ in a given $\zz$ renormalization transformation $R$. To understand this behavior, let us compare the RG transforms $BD$ and $DB$ acting on a 2D complex spin model. Writing $z$ in polar form as $re^{i\theta}$, we find explicitly that
\begin{align}\label{eq:bd-phase-factor}
	BD(e^{i\theta} r)={2r^2e^{i2\theta}+r^4 e^{-4i\theta}\over 1+2r^4}\,.
\end{align}
\Eq{bd-phase-factor} only contains a single overall phase factor if $\theta=\pi n /3$: along these constant-$\theta$ lines, $|z|$ evolves just as it would along the positive real axis. Off these lines, however, the phase can evolve, thus explaining the six-fold symmetry of the $BD$ phase diagram. On the other hand, 
\begin{align}\label{eq:db-phase-factor}
	DB(e^{i\theta} r)=\left({2r e^{i\theta}+e^{-2i\theta}r^2 \over 1+2r^2}\right)^2\,.
\end{align}
\Eq{db-phase-factor} has an overall phase if $\theta = 2\pi n/3$, giving the phase diagram threefold symmetry. 

If we now consider the general Migdal-Kadanoff transform for a complex $\zz$ spin model
\begin{align}
	R(z)=D^{d-1-j}BD^j(z)\,,
\end{align}
we see that 
\begin{align}
	D^j(e^{i\theta}r)=e^{2ji\theta}D^j(r)
\end{align}
suggesting a proliferation of copies of the positive real axis with increasing $j$. However, $B(e^{i\theta} r)$ produces copies of the positive real axis only for $\theta = 2\pi n/3$. Subsequent application of $D$ does not produce further copies of the positive real axis, so the phase diagram has a $3\times 2^j$-fold symmetry, in agreement with the observed behavior. The analysis of other cases follows from similar reasoning. 

We see that the real-space RG cannot answer which symmetry is correct on a standard cubic lattice for complex and chiral models.  All of the models have a $\zz$-symmetric action, with no indication that a larger-symmetry group is somehow hidden in the action. This logic favors the RG scheme that produces a three-fold symmetric phase diagram as the correct one. Another argument in favor of the three-fold symmetric phase diagram, which relies on properties of chiral model low-temperature expansions, is given in Appendix \ref{app:config-worldline}.

\subsection{Presence and absence of Devil's flowers}

In $d\geq 3$, only chiral spin models and their complex duals  have a Devil's flower phase structure. The Devil's flower appears for all possible RG scheme choices $R$ in these models, though the size of each phase and the symmetries differ based on the scheme. 

One might wonder whether the Devil's flower stems from Roberge-Weiss symmetry, as the chiral spin models and their duals are unique among our models in having $p=2$. As noted in \cite{PhysRevB.24.5180}, for a blocking factor $\lambda=2$, we have one fixed point $z=1$ and a two-cycle $z=\{\Omega,\Omega^2\}$; thus, it makes sense that adjacent phases of the Devil's flower map after many iterations to different points in $\{1,\Omega,\Omega^2\}$. However, for $\lambda=3$ we have $\tilde B_3(\omega z) = \omega \tilde B_3(z)$ and $\tilde D_3(\omega z) =  \tilde D(z)$. For a chiral spin model, $R_3(z)$ contains only one $\tilde D_3$, implying that $p=0$. However, the lines separating the different regions still map into all three elements of $\{1,\Omega,\Omega^2\}$. In other words, the phase structure is not gone for $\lambda =3$, but is just harder to discern. That is, $p=2$ is not required for a Devil's flower. 

Instead, let us write the coarse-grained order parameter for chiral spin models as $M(x) = \rho(x) \exp[i\phi(x)]$. Symmetry arguments suggest the Landau free energy density takes the form 
\begin{align}\label{eq:fk1}
	f={1\over 2}(\nabla+iA) M^* \cdot (\nabla-iA) M +BM^*M+C(M^3+M^{*3})+\cO(M^4)
\end{align}
where $A$ is a constant vector field in the chiral direction, and $\{B,C\}$ are scalar parameters. In the low-temperature limit, we can take the magnitude of $M$ and $\rho_0$ to be fixed and reduce the Landau free energy density to
\begin{align}\label{eq:fk2}
	f={1\over 2}\rho_0^2\left(\nabla \phi - A\right)^2-2 C\rho_0^3 \cos(3\phi).
\end{align}
Minimization of \eq{fk2} leads to the fundamental equation of the \textit{Frenkel-Kontorova model} \cite{FK_model}, a well-known system with Devil's staircase behavior. (The explicit form of the parameters in the model can be easily obtained using, for example, mean field theory \cite{ottinger_two-dimensional_1983}.) The potential term favors a constant value of $\phi = 2\pi n/3$, while the kinetic term is minimized if $\phi$ changes at a constant rate along chiral direction. In the Frenkel-Kontorova model, the interplay between these two terms leads to a Devil's staircase for $\phi$ along the direction of $A$. For general $\zn$ models, \eq{fk2} generalizes to
\begin{align}
	f={1\over 2}\rho_0^2\left(\nabla \phi - A\right)^2-2 C\rho_0^N \cos(N\phi)\,.
\end{align}
The key feature of this connection is the presence of a local order parameter $M$ in the chiral spin models and their duals. However, for the other cases, the fundamental field of the chiral model is not an order parameter due to \textit{Elitzur's theorem} \cite{Elitzur:1975im}, which prohibits the spontaneous breaking of local gauge symmetry in gauge theories without gauge fixing. Generally, unless an Abelian lattice model or its chiral dual has a local order parameter, it cannot support a Devil's flower.

\subsection{Extension to finite temperature}

So far we have only included the effects of a nonzero chemical potential in an infinite 
Euclidean spacetime. 
To build models with a closer relationship to QCD at nonzero density and temperature, we must also consider the effects of nonzero temperature, obtained by giving the lattice a toroidal topology, $\mathbb{R}^3\otimes \mathbb{T}^1$, where the circumference of $\mathbb{T}^1$ is $\beta=1/T$, the inverse temperature. 
Note that the temperature introduced here is completely distinct from our previous use of low- and high-temperature to encapsulate our intuition about phases for $J\gg 1$ and $J \ll 1$. Here, we take the direction associated with temperature to be that of $\theta$ (chemical potential); we refer to this as the Euclidean time direction.

In general, when $\beta$ is much larger than any correlation length in a system, the system's behavior resembles that of $\beta = \infty$. When we decrease $\beta$ to become commensurate with one of the correlation lengths, finite temperature effects become non-negligible. From an RG perspective, the RG flow near a $d$-dimensional fixed point typically crosses over to behavior associated with a $(d-1)$-dimensional fixed point. This is closely associated with \textit{dimensional reduction}, the description of a $d$-dimensional field theory at high temperature by an effective $(d-1)$-dimensional theory. This behavior occurs naturally in the Migdal-Kadanoff RG. For a complex spin model in $d$ dimensions, the RG transformation is
\begin{equation}
	R_d(z)=DB^{d-1}(z)
\end{equation}
for all directions.
When $\beta=2$ in lattice units, periodic boundary conditions in time imply that there is no spin-spin interaction in that direction anymore; it has been integrated out. This leads to a dimensionally-reduced spin model in $(d-1)$ dimensions with no chemical potential. The RG flow for this reduced model evolves as
\begin{equation}
	R_{d-1}(z)=DB^{d-2}(z).
\end{equation}

A more interesting behavior occurs for gauge theories. The Polyakov loop $P$,
a closed thermal Wilson line,
is the product of the link variables $U(\vec x, t)$ winding around the lattice in the time direction:
\begin{equation}
	P(\vec x, t)=U(\vec x, t)U(\vec x, t+1)\dots U(\vec x, t-1)
\end{equation}
where periodic boundary conditions close the loop. Note that $\Tr \, P(\vec x, t)$ is gauge-invariant and independent of the choice of $t$, and it functions like a scalar from a $(d-1)$-dimensional perspective. This is the order parameter for the deconfinement phase transition in a pure gauge theory at nonzero temperature.
Using, for example, the original form of the  Migdal-Kadanoff RG, the complex $\zz$ gauge theory has the RG transformation
\begin{equation}
	R_d(z)=D^2 B^{d-2}(z).
\end{equation}
When we repeatedly act with $R$ on timelike plaquettes, we eventually reach a point where the system is effectively $(d-1)$-dimensional, like before. However, the timelike gauge interactions remain between neighboring Polyakov loops $P(\vec x, 0)$, effectively generating a spin-spin interaction between them.\footnote{This is a concrete realization of \textit{Svetitsky-Yaffe universality} \cite{Svetitsky:1982gs}: the deconfinement transition of a $d$-dimensional gauge theory lies in the same universality class as a $(d-1)$-dimensional spin system using $\Tr\,P(\vec x,t)$ as spin variables.} In the Migdal-Kadanoff framework, once we reach the limiting case of a $(d-1)$-dimensional spin system, the timelike couplings evolve as
\begin{equation}
	R_{d-1}(z)=D B^{d-2}(z)
\end{equation}
which is precisely the evolution of a $(d-1)$-dimensional spin system. At this point, a $d$-dimensional $\zz$ gauge theory evolves as two separate systems: a $(d-1)$-dimensional $\zz$ spin system with Polyakov loops playing the roles of spins and a $(d-1)$-dimensional $\zz$ gauge theory. This decoupling is unique to Abelian gauge theories; for non-Abelian gauge theories, the Polyakov loops act as adjoint scalars in $(d-1)$ dimensions and remain coupled to the gauge field. In the case of chiral $\zz$ gauge theory, the dimensionally-reduced $\zz$ spin system is chiral, and the gauge theory is Hermitian. For a complex $\zz$ gauge theory, the dimensionally-reduced $\zz$ spin system is complex. In four dimensions, a chiral $\zz$ gauge theory at nonzero temperature has a Devil's flower phase structure associated with Polyakov loops, but the corresponding complex $\zz$ gauge theory does not.

\subsection{Extension to $\zn$}

Carrying out real-space RG analyses of $\zn$ models becomes increasingly more complicated for increasing $N$. To parametrize nearest-neighbor $\zz$ models, we only need one parameter in the Hermitian case and two parameters in the chiral and complex cases. As we increase $N$, the number of required parameters increases according to the number of nontrivial simple representations of the group. For example, $\bZ_4$ has 4 irreps, associated with the mapping $s \to \{1, s, s^2, s^3\}$, two of which are complex and conjugate ($s^3=s^*$) and one of which is real ($s^2$); thus, it requires one complex and one real parameter. 

While we do not work out the behavior of $\zn$ models explicitly, it seems likely that phase diagram characteristics of $\zz$ models will persist for all $N$. For example, chiral $\cZ_4$ spin models are known to exhibit Devil's flowers for $d\geq 3$ \cite{Yeomans1982}; we expect chiral spin models and their duals to exhibit this structure for all $N$ in $d\geq 3$. Likewise, we expect that $\zn$ complex spin models have similar phase diagrams as the $\zz$ case, with a disordered phase and at least $N$ ordered phases; however, as discussed above, we expect that the correct $\zn$ result for cubic lattices is a disordered phase and \textit{exactly} $N$ ordered phases. 

\subsection{Extension to SU($3$)}
While neither $\cZ_2$ nor SU(2) models have complex irreducible representations (irreps), extending our $\zn$ results to complex and chiral SU($N$) models for $N \ge 3$ is more complex because SU($N$) has an infinite number of irreps. Nonetheless, Migdal-Kadanoff methods have been applied to Hermitian SU($N$) gauge theories in 4D and $d=4-\epsilon$ \cite{Caracciolo:1978wr, Menotti:1981ry}, which provides us a starting point. 

As with $\zn$ gauge theories, 2D gauge models are exactly solvable. The Migdal-Kadanoff RG consists only of decimations, which map a class of  SU$(N)$ heat kernel actions, parametrized by a single real parameter $\beta$, into itself. These actions give results essentially identical to those of continuum 2D SU($N$) pure gauge theories. There are two fixed points: an unstable high-temperature fixed point at $\beta=\infty$ and a stable low-temperature fixed point at $\beta=0$, similar to the behavior of a standard 1D $\zz$ spin model. In 4D, a perturbative analysis around $\beta=\infty$ shows that this is a good approximation to the standard UV fixed point at $g^2=0$. In $d=4+\epsilon$, that fixed point becomes an IR fixed point, and a new UV fixed point at $g^2={\mathcal O}(\epsilon)$ emerges. This is consistent with the continuum behavior of such theories; parallel results hold for SU($N$) spin models in $d=2+\epsilon$ \cite{Brezin:1975sq, Friedan:1980jf}.

Lattice SU($N$) models display additional structure when we extend the usual Wilson lattice action to include a term in the adjoint representation. For example, for a gauge theory the extended action is \cite{Bhanot:1981eb, Bitar:1982bp}
\begin{align}
	S[U_p]=\sum_p \left[{\beta_F \over 2 N}\left(\chi_F(U_p)+\chi_F(U_p)^*\right)+
	{\beta_A \over N^2-1}\chi_A(U_p)\right].
\end{align}
Here $\chi_R$ is the group character for the irrep $R$: $\chi_F(U_p)$ is the trace of the plaquette variable $U_p$ in the fundamental representation, and $\chi_A(U_p)=\left|\chi_F(U_p)\right|^2-1.$
In the limit $\beta_A\rightarrow\infty$, the plaquette variables take on values in $\zn$. Thus in this limit, an SU($N$) gauge theory reduces to a $\zn$ gauge theory, and similarly for SU($N$) spin systems.

We can also define chiral and complex SU($N$) lattice models; for spin models, we have
\begin{align}\label{eq:sun_models}
	S_{c}[U_x]&=\sum_{x,r}\Big[\frac{\beta_F }{2N}\left(\chi_F(U_{x+\hat r}^+ U_{x})+ \chi_F(U_{x-\hat r} U_{x}^+)\right)+\frac{\theta_r}{N\sqrt{3}}\left(\chi_F(U_{x+\hat r}^+ U_{x})- \chi_F(U_{x-\hat r} U_{x}^+)\right)
	\nonumber\\
	&\qquad\qquad+\frac{\beta_A }{N^2-1}\chi_A(U_{x+\hat r}^+ U_{x})\Big]
	\\
	S_{\chi}[U_x]&=\sum_{x,r}\left[\frac{\beta_F }{2N}\left(e^{i\tilde\theta_r} \chi_F(U_{x+\hat r}^+ U_{x})+e^{-i\tilde\theta_r} \chi_F(U_{x-\hat r} U_{x}^+)\right)
	+\frac{\beta_A }{N^2-1}\chi_A(U_{x+\hat r}^+ U_{x})\right]\,,
	\nonumber
\end{align}
where $\{x,r\}$ index lattice sites and directions, and $\{\theta_r,\tilde \theta_r\}$ are nonzero only along the complex or chiral direction, respectively. The normalizations of the parameters agree with the $\zz$ models in the limit $\beta_A \to \infty$.

There is no known general connection between the two models in \eq{sun_models} except in the $\zn$ limit. However, the first-order transitions we have found for $\zz$ models should extend to large but finite values of $\beta_F$, which allows us to make predictions for $T=0$ behavior. We expect that chiral SU($N$) spin models in $d\geq 3$ exhibit a Devil's flower phase structure as $\tilde \theta$ is varied for at sufficiently large $\beta_F$ and $\beta_A$. We also expect that chiral SU($N$) gauge models for $d\geq 3$ exhibit a four-phase structure as we vary $\tilde \theta$ at sufficiently large $\beta_F$ and $\beta_A$. Similarly, we expect that complex spin and gauge models in $d\geq 3$ exhibit a four-phase structure as we vary $\theta$ for small $\beta_F$ and large $\beta_A$. 

\section{Outlook}\label{sec:outlook}

In this work, we studied lattice $\zz$ spin and gauge theories with complex and chiral interactions. We demonstrated an extension of Kramers-Wannier duality that maps complex and chiral models onto one another. We analyzed the phase structure of these models with real-space RG methods, and showed how $\zz$ and $\cCK$ symmetries impact the RG flow and phase structure. In particular, we showed that spatially-modulated phases appear in both chiral and complex models, and that these phases manifest in a Devil's flower structure in the specific case of chiral spin models and their duals. 

We have shown that the phase structure of $\zz$ non-Hermitian models obtained from Migdal-Kadanoff RG calculations depends strongly on the order of bond moving and decimation, a violation of expectations of universality. This finding underscores the need to explore whether other real-space RG formulations, and other theoretical tools like mean field theory exhibit similar issues when applied to non-Hermitian and finite-density models.
In cases where a real dual form is available, standard lattice methods can in principle determine the correct phase structure. In practice, it may be necessary to consider both order and disorder variables for a complete understanding.

We have also found that the Devil's flower phase structure only occurs in a given model if there is a gauge-invariant order parameter available in the model or its dual, a behavior we trace to Elitzur's theorem. This has profound implications for the appearance of inhomogeneous phases in lattice field theories in general.

We have partially extended our results on $\zz$ to $\zn$ and SU($N$) lattice models. The $\zn$ chiral spin models with $N\ge 3$ are all expected to have $\zn$ spirals. As in the $\zz$ case, we predict that chiral SU($N$) spin models exhibit a Devil's flower phase structure for $d\geq3$, but that chiral gauge theories do not, due to Elitzur's theorem. Unfortunately, the lack of a simple duality for non-Abelian lattice models prevents us from making corresponding statements about complex SU($N$) lattice models, which are more directly related to finite-density QCD. It is possible that chiral spirals and $\zn$ spirals may have a synergistic effect for models in which Polyakov loops and fermion bilinear order parameters are coupled, such as PNJL models.

\section*{Acknowledgments}
We thank Zohar Nussinov and Jesse Thaler for helpful discussions.
S.T.S. was supported by the U.S. Department of Energy, Office of Science, Office of Nuclear Physics from DE-SC0011090; the U.S. National Science Foundation through a Graduate Research Fellowship under Grant No. 1745302; fellowships from the MIT Physics Department and School of Science; and the Hoffman Distinguished Postdoctoral Fellowship through the LDRD Program of Los Alamos National Laboratory under Project 20240786PRD1. Los Alamos National Laboratory is operated by Triad National Security, LLC, for the National Nuclear Security Administration of the U.S. Department of Energy (Contract Nr. 892332188CNA000001). 

\appendix
\section{Configuration-worldline duality}\label{app:config-worldline}

Chiral $\zz$ spin models have an unusual low-temperature (small $\tilde J$) phase structure for $d\ge 3$, with an infinite number of periodic inhomogeneous phases commensurate with the underlying lattice. Evidence for this behavior was first obtained via low-temperature ($T$) expansions \cite{Yeomans1981}, and has also been indicated by other approaches \cite{PhysRevB.24.398,PhysRevB.24.5180,selke_monte_1982,mccullough_mean-field_1992}.  
In a given inhomogeneous phase, $(d-1)$-dimensional sheets of spins, each characterized by a certain expectation value, are layered along the chiral direction, forming a $\zz$ spiral along the chiral direction. 

A low-$T$ expansion of a spin system typically entails expanding about the lowest-energy configuration(s); in our sign convention, these configurations have the largest lattice action. In our $\zz$ chiral spin models at $\tilde\theta = 0$, the ground state configurations have all their spins aligned, leading to three possibilities for the starting point for our expansion (one for each possible spin). Higher-order terms in the expansion reflect contributions to the partition function from configurations formed by flipping one or more spins in the original configuration. 

As we vary $\tilde\theta$, configurations become inhomogeneous in the chiral direction $\tilde\theta$ but remain homogeneous in the directions transverse to $\tilde\theta$; this behavior is called a chiral spiral. 
We describe these configurations by the element of $\zz$ associated with each transverse slice of the lattice configuration, using $n=0,1,2$ as a shorthand to denote the three $\zz$ spins $\exp(2\pi n/3)$. 
For example, we can write the three homogeneous ground states at $\tilde\theta = 0$ as a sequence of period one: $(000...)$, $(111...)$, and $(222...)$. These configurations are all equivalent up to a global $\zz$ rotation.
Likewise, at the special values $\tilde\theta = 2\pi/3, 4\pi/3$, the ground state configurations have period three; specifically, repeated sequences of $(012)$ and $(021)$, up to a global rotation. 
Every value of $\tilde\theta$ is associated with a sequence of spins, each associated with a phase. 
A first-order transition occurs when the free energies of two stable phases are degenerate in parameter space.

An alternative notation for describing field configurations tells us how many times a spin value repeats before changing. For example, in the configuration $(012)$, the spin value repeats once before changing, which we can write as
\begin{equation}
	(0\,1\,2\,0\,1\,2\,...\,)  =\langle 1 \rangle\,.
\end{equation}
We can generalize this notation to more complicated patterns $\langle k_1 \,k_2\, ... \,k_n\rangle$. For example, we could have a spin configuration that looks like 
\begin{align}
	\{0\,11\,22\,000\,1\,22\,00\,111\,2...\} = \langle 12^23\rangle\,.
\end{align}
The righthand side indicates that there is a layer consisting of one spin, followed by a layer of two spins, followed by another layer of two spins, followed by a layer of three spins, before the pattern repeats. The notation
\begin{align}
	\{000...\} = \{111...\} = \{222...\} = \langle \infty \rangle 
\end{align}
indicates a homogeneous phase. 

Due to the duality between the low-$T$ expansion of a chiral model and the high-$T$ expansion of its corresponding complex model, we can associate configurations of chiral models with contributions to the high-$T$ expansion of complex models. For example, in $d=2$ spin systems, duality relates each spin configuration of the chiral model to a set of closed paths in a worldline expansion of the corresponding complex model, and vice versa. In $d=3$, spin configurations of the chiral spin model are dual to terms in a worldsheet expansion of the dual complex gauge theory. We refer to this relationship as spin configuration-worldline duality.
\begin{table}
	\begin{center}
		\begin{tabular}{|c|c|c|c|c|}
			\hline
			Brief notation & Spin configuration & Worldlines 
			\\\hline
			$\langle \infty \rangle $&$(000000)$ & $[000000]$ 
			\\\hline
			$\langle 3\rangle $&$(000111222)$ & $[001001001]$ 
			\\\hline
			$\langle 32\rangle $&$(000112220011122)$ & $[001010010100101]$ 
			\\\hline
			$\langle 32^2\rangle $&$(0001122)$ & $[0010101]$ 
			\\\hline
			$\langle 2\rangle $&$(001122)$ & $[010101]$ 
			\\\hline
			$\langle 1\rangle $&$(012012)$ & $[111111]$ 
			\\\hline
		\end{tabular}
	\end{center}
	\caption{Examples of spin configuration-worldline duality.} \label{table:spins-and-worldlines}
\end{table}

In $d=2$, configurations of the chiral spin model are mapped by duality into straight worldlines in the direction transverse to the chiral direction. 
In between each two adjacent slices of the $(012)$ configuration, there is a worldline carrying $\zz$ triality $1$. Similarly, in between each two adjacent slices of the $(021)$ configuration, there is a worldline carrying $\zz$ triality $2\equiv -1$. We can associate the spin configurations described above with their corresponding worldlines as follows:
$$(000)\leftrightarrow [000]$$
$$(012)\leftrightarrow [111]$$
$$(021)\leftrightarrow [222]$$
The worldline value on the right is the difference of the successive spins on the left.
We can think of $[111]$ and $[222]$ in $d=2$ as representing constant currents in the Euclidean timelike direction, arising from a nonzero chemical potential.
We give further examples of this configuration-worldline duality in Table \ref{table:spins-and-worldlines}.

The spin configurations $(012)$ and $(021)$ are atypical in the chiral models: they are inhomogeneous, but their worldline duals $[111]$ and $[222]$ are homogeneous. 
From this analysis, we see that it is natural that a $\zz$ model without a Devil's flower has only four phases, with a three-fold symmetric phase diagram: Each of the three ordered phases of a chiral model is obtained in the complex form of the model from a starting point of constant density.

\bibliographystyle{JHEP}
\bibliography{v3}

\end{document}